# Valuing Insurance Against Small probability risks: A Meta-Analysis

Selim Mankaï[†]    Sébastien Marchand[∓]    Ngoc Ha Le[‡]

University of Clermont Auvergne

November 24, 2023

**Highlights:**

- Average stated willingness to pay (WTP) for insurance is less than the expected losses.
- Probability information provision increases average stated WTP.
- Cultural factors explain differences in insurance demand across countries.
- Data collection design and sampling influence elicited WTP across studies.
- Need to consider the interactions between fundamental and methodological factors on WTP.

**Abstract:**

The demand for voluntary insurance against low-probability, high-impact risks is lower than expected. To assess the magnitude of the demand, we conduct a meta-analysis of contingent valuation studies using a dataset of experimentally elicited and survey-based estimates. We find that the average stated willingness to pay (WTP) for insurance is 87% of expected losses. We perform a meta-regression analysis to examine the heterogeneity in aggregate WTP across these studies. The meta-regression reveals that information about loss probability and probability levels positively influence relative willingness to pay, whereas respondents' average income and age have a negative effect. Moreover, we identify cultural sub-factors, such as power distance and uncertainty avoidance, that provided additional explanations for differences in WTP across international samples. Methodological factors related to the sampling and data collection process significantly influence the stated WTP. Our results, robust to model specification and publication bias, are relevant to current debates on stated preferences for low-probability risks management.

[†] Department of Finance, IAE Clermont Auvergne-School of Management, CleRMa, France (Corresponding author: selim.mankai@uca.fr);

[∓] Department of Economics, CERDI, University of Clermont Auvergne, France

[‡] Ph.D. candidate, CERDI, University of Clermont Auvergne, France.

This work was supported by the International center of research (CIR4), [Grant: 2019CYCLONE-CIR4].

# 1. Introduction

Low-probability, high-impact (LPHI) adverse events challenge our understanding of the interplay between individuals' risk perception and risk-coping behaviors (Kunreuther et al., 2013, 2021). Natural disasters and extreme weather events, as a salient case, can have a significant cost in terms of human lives and physical damages. These events exert substantial pressure on both individuals and society and exacerbate existing inequalities, particularly when their impacts are unevenly distributed (Klomp and Valckx, 2014, Botzen et al., 2020). To mitigate the economic impacts of natural disasters, policymakers provide financial assistance and promote insurance programs as a financing mechanism to smooth individuals' consumption against income shocks.[2] Paradoxically, the demand for non-mandatory insurance against low-probability risks is relatively low even when it is provided at prices below actuarially fair values, (Baillon et al. 2022).

Conventional models based on expected utility theory (EUT) posit that the main factor influencing the insurance decision is the agents' attitude towards risk (Arrow, 1971). However, numerous observations show that insurance take-up is much lower for events with low-probability and high-consequences than for those with higher probability and lower consequences. From an EUT perspective, this evidence therefore seems puzzling, as it calls into question the rationality hypothesis (Sydnor, 2010). Behavioral economics experiments show that people do not always make decisions consistent with the normative EUT benchmark, which casts doubt on the adequacy of this model to explain the low demand for LPHI insurance. Prospect theory (PT), rank-dependent utility (RDU), and cumulative prospect theory (CPT), formulated respectively by Kahneman and Tversky (1979), Quiggin (1982), and Tversky and Kahneman (1992), provide the most promising alternative to the EUT model. These descriptive models incorporate additional components of risk preferences, such as reference-dependent behavior, probability weighting, and loss aversion.

One particular facet of the demand disparity between LPHI and HPLI– the difficulty in understanding small probabilities – has become the focus of intense investigation over the past years. Some research even goes further, suggesting that people often confuse low-probability with zero-probability events (Kunreuther et al. 2001). Due to the complexity of analyzing low-probability events in real-world settings and the inherent challenges in establishing causal effects for fundamental factors, research in the past two decades has increasingly turned towards stated preference methods and willingness to pay (WTP) estimation. These methods offer a more flexible framework for measuring the demand curve, taking various conditions into account, providing further evidence of the "demand puzzle" (Jaspersen, 2016; Robinson and Botzen, 2019).

Additional explanations using stated preference techniques have been proposed from the psychological and behavioral economic literature to explain why individuals underestimate these risks (Friedl et al., 2014, Browne et al., 2015, Fehr-Duda and Fehr, 2016). Low demand for insurance can be caused by search costs associated with gathering information about insurance premiums, coverage, and underlying probabilities of loss (Kunreuther and Pauly, 2004). Lack of loss experience has also been shown to be consistently related to low insurance demand, which is perhaps driven by low perceptions of risk (Robinson and Botzen, 2019). It has been shown that these methods of communicating risk can have a positive effect on demand for risk reduction and can increase the sensitivity of demand to probability changes (Botzen and Van Den Bergh, 2012). Similarly, the willingness to pay for insurance is significantly lower for correlated than for idiosyncratic risks. Friedl et al. (2014) claim that insurance is less attractive for correlated risks, a premise that they theoretically exposed in a model with a social reference point.

Risk perception for small probability events is subject to cognitive biases and judgment-distorting heuristics (Botzen and Bergh 2009; Gallagher 2014). A relatively recent strand of literature related to stated preferences focuses on these behavioral frictions as local drivers of underinsurance. Pitthan and De Witte (2021) summarize the moderating effect of these factors on the relationship between extreme risk attributes, risk perception, and insurance demand.[3]

---

[2] By providing a timely post-loss financial payout, insurance is also linked to more positive psychosocial and emotional outcomes for affected people (Farrell and Greig 2018; McKnight 2019). The insurance industry may similarly support public risk management policies by promoting preventive measures and by providing relevant information to the public. Hudson et al. (2017) show that insured populations are more likely to undertake disaster preparations and preventive actions than the non-insured.

Despite these non-market data methods' effectiveness in measuring the demand curve, the literature documents, however, high variability in stated WTP.[4] Experience-based studies provide relatively high average WTP compared to expected loss, pointing to significant upward bias attributable to the data collection method (Leblois et al., 2020).[5] For instance, Zimmer et al. (2018) conduct a laboratory experiment to examine probabilistic insurance demand for low-probability risks using real rewards applied in an incentive-compatible framework. In the absence of default risk, the average WTP for insurance is higher than expected losses. In framed field experiments conducted with farmers, Serfilippi et al. (2020) analyze the micro-insurance demand to explain low take-up rates. The average WTP for a virtual insurance contract under a standard frame also exceeds the expected loss. In the same vein, Robinson et Botzen (2019) explore the flood insurance demand through an online experiment conducted with a sample of Dutch homeowners. They find that individuals inclined to purchase insurance are willing to pay a significant mark-up over expected losses.

Some factors related to data collection methods and elicitation mechanisms can accentuate hypothetical biases or strategic behavior (e.g. Balistreri et al., 2001; Harrison and Rutström, 2008, Entem et al., 2022). This variability is further exacerbated by the high degree of heterogeneity between studies conducted in different contexts (e.g., periods, regions, risk types, insured assets, etc.). All these factors may limit the ability to answer some relevant questions about the true elicited insurance demand for low-probability risks, including: 1) How can all the estimated average WTP results reported in the literature be aggregated? 2) Is there publication bias in the literature? 3) How low is the stated demand for insurance?

While the focus has been on local factors affecting demand at the individual level, little attention has been paid to assessing, at the study level, the cumulative effects of risk attributes, socio-economic characteristics, and methodological factors on the gap between expected loss and average WTP estimates. As the number of studies on insurance demand against low portability risks is expected to keep growing in the coming years, particularly with climate change, disentangling fundamental and measurement effects on aggregate stated WTP estimates is highly needed.

In this paper, we present a meta-analysis of contingent valuation studies focusing on low-probability risk insurance.[6] The main goal of our study is to explain variations in the average stated willingness to pay (WTP) across studies, rather than delving into the individual factors influencing the demand for insurance. Although these two objectives may initially appear similar, they hold distinct significance within our research context. We specifically rely on available aggregated data, recognizing that factors that exert influence at the individual level may not hold the same relevance at the aggregate level (Schmid et al., 2020). Moreover, we introduce new meta-factors, such as data collection strategies and survey designs, which may not directly impact individually stated WTP.

---

[3] For instance, Kunreuther (2021) describes six cognitive biases in the context of the US flood insurance market: myopia, amnesia, optimism, inertia, simplification, and herding. Another example of behavior anomaly, to name but one, assumes a specific type of narrow framing that views insurance as an investment, underestimating its primary purpose as a way to manage risks and protect against unexpected extreme losses (Platteau et al., 2017).

[4] Contingent valuation methods may suffer from several limitations attributed mainly to the hypothetical nature of the survey that tends to overestimate the true willingness to pay. For example, individuals might not be able to judge the value of the goods they have to evaluate, due to a lack of understanding; or they might have difficulty envisioning their income constraints in the proposed hypothetical setting (Diamond et al., 1994). These limitations may dramatically reduce external validity (Haghani et al., 2021).

[5] Leblois et al. (2020) note a high take-up of insurance giving as examples the following studies (Petraud et al., 2015; Norton et al., 2014; Serfilippi et al., 2020). They explain such discrepancy by the presence of seasonal liquidity constraints and distrust in insurance providers.

[6] Our analysis is specifically designed to study insurance-related decision-making about material losses and natural hazard damages, deliberately excluding health risks. Although health insurance is an important area of research in its own right, it is governed by distinct regulatory frameworks and involves unique risk assessment methodologies and psychological influences on decision-making behavior.

Based on a sample of 38 primary studies (65 observations after removing outliers) experimentally elicited and survey-based measures spanning 17 years of research, we find a meta-analytic average WTP, before heterogeneity treatment of 87% of expected losses. WTP estimates vary considerably across observational and experimental studies. To explore the sources of heterogeneity, we perform a meta-regression analysis (MRA hereafter) controlling for a range of moderators and applying several estimation methods. As a robustness check, we use the Bayesian Model Averaging approach (BMA) to account for model uncertainty and covariate selection.[7]

Our main finding is related to the conditions under which the willingness to pay for insurance may deviate from trend values. Moderators such as information about loss probability provision and very small probability levels influence positively relative WTP, whereas respondents' socio-demographic characteristics such as income and age have a negative effect. Laboratory-based estimates, show-up fees, and within-subject design appear to report significantly higher values for relative WTP than other methods. We identify other interesting determinants that affect the relative WTP through different causal channels. None of the two moderators linked to this aspect (i.e. incentive-based elicitation and questions format) is robustly significant. Our results provide no direct support for hypothetical bias traditionally associated with contingent valuation methods.

The relative WTP estimates seem to be geographically dependent, showing smaller levels in China compared to Germany and the Netherlands. This finding is highly consistent with previous empirical studies conducted in developing countries. To gain insight into these results and search for a possible explanation of large differences across these countries, we explore the impact of national culture using Hofstede's cultural dimensions. We find that relative WTP is positively (negatively) related to "Uncertainty avoidance" ("Power distance") dimensions. Culture proxies thus provide statistically robust drivers of insurance demand supporting previous results of Chui and Kwok (2008) and Park and Lemaire (2012). We perform various robustness checks to assess the sensitivity of the results to various modeling choices. Our main results are robust to alternative estimation specifications and publication bias issues.

To our knowledge, this meta-analysis is the first to focus on the underinsurance issue from this angle and extends, with a new quantitative perspective, previous surveys of literature (e.g. Jaspersen, 2016; Robinson and Botzen, 2019; Harrison et al., 2019; Lucas, 2021). The current study provides new results on the effect of fundamental and methodological factors on WTP. It also allows for testing the literature for potential publication bias, a goal that cannot be achieved at the individual study level (Stanley et al., 2012). Additionally, it allows us to revisit how different cultural dimensions may have distinct conceptions of resilience at the organizational or community level that may persistently influence insurance demand. Finally, this study contributes to examining the presence of hypothetical bias for various WTP elicitation designs (Lusk and Schroeder, 2004; Miller et al., 2011; Schmidt and Bijmolt, 2020).

The rest of the paper is structured as follows. Section 2 presents briefly the general theoretical framework underpinning the willingness to pay for insurance. Section 3 describes the meta-dataset and discusses the metric construction. Section 4 examines publication bias. Section 5 investigates heterogeneity and presents the MRA results. In this section, we further check the robustness of the obtained results. The last section concludes.

## 2. Theoretical background

As a conventional measure of the change in welfare, compensating variation is defined as the maximum amount an individual would be willing to pay (WTP) to secure a change (i.e. restore the original welfare level) (Hanemann, 1991). For insurance contracts without deductibles or other cost-sharing limits, the willingness to pay (WTP) is implied from the following agent indifference condition between insurance and non-insurance decisions:

$$U(y - WTP, p, q_1; Z) = U(y, p, q_0; Z) \quad (1)$$

where $U$ describes the agent's indirect utility (value) function[8], $y$ represents the individual's wealth (income), $p$ is a vector of costs that the individual faces, $q_i$ (with $i=0, 1$ and $q_0 < q_1$) reflects the safety value, and $Z$ is a vector of personal characteristics (e.g. past loss experience, financial literacy age, gender, etc.). Parameters $q_0$ and $q_1$ describe different levels of the safety measure $q_i$. Parameter $q_1$ is associated with a measure that provides a higher level of safety compared with $q_0$ (Entorf and Jensen, 2020).

---

[7] Our empirical analyses are performed according to meta-analysis guidelines (e.g. Havranek et al., 2020; Steel et al., 2021).

Differences in insurance demand at the individual level could be attributed to several fundamental factors, which can be derived from various risk preference models. According to the static (EUT) framework, WTP would equal the certainty equivalent of the expected utility over final wealth states without insurance. The difference between the fair premium and the theoretical WTP corresponds to the standard risk premium, which is a function of the risk aversion level and the probability distribution of losses.[9] However, in practice, the disparity between EUT predictions and stated WTP calls for more extra risk premiums. Baillon et al., (2022) introduce a behavioral decomposition of the gap between stated willingness to pay (WTP) and fair insurance price inspired by the prospect theory (PT) model:

$$WTP - \mu = \pi^b + \pi^u + \pi^w + \pi^\varepsilon \quad (2)$$

They consider three behavioral deviations from the fair price $\mu$ arising from subjective beliefs $\pi^b$, convex utility in the loss domain $\pi^u$, and probability weighting $\pi^w$. A residual term $\pi^\varepsilon$ is also considered to absorb all influences on WTP that are not captured by the model. In both EUT and PT models, the loss probability plays a major role in insurance decisions. In practice, two distinct behaviors may be observed depending on the probability level. When people deem the probability of loss below their level of concern, they generally neglect risk and choose not to undertake protective actions (Kunreuther and Pauly, 2004; Kunreuther et al., 1978). In contrast, when they attribute a subjective likelihood of loss that is far higher than the actual probability, they become highly risk-aware and seek out mitigation strategies (e.g., Brouwer et al., 2014). Furthermore, Abito and Salant (2019) find that the provision of information on a loss probability reduces both the subjective probability and WTP. Other models, based on the dynamic consumption utility framework, posit that WTP primarily depends on expected losses, wealth, annual income, market interest rates, and risk attitudes (Hansen et al., 2016). As the agent's risk preferences are unobservable and challenging to measure objectively, it is of relevance to get insight into observable factors that moderate risk attitude determinants and structurally impact demand. These factors can be globally grouped into three categories: demand-side, supply-side, and extreme risk nature (Leblois et al., 2020).

In line with the goal of the study, we rely solely on available aggregated data related to study-level covariates. This is crucial given the inherent complexities in determining an agent's true WTP. Since WTP is unobservable and there is no widely acknowledged elicitation method (Völckner, 2006), methodological choices may influence final results. Carson et al. (2001) give some guidance in this area and outline two important aspects of the elicitation methodology. The first consideration is whether the elicitation process is based on price generation (i.e., an open-ended question format) or price selection tasks (i.e., a dichotomous choice question format) (Hofstetter et al., 2021). The dichotomous choice typically overstates WTP relative to open-ended (Balistreri et al., 2001) and payment card formats (Ready et al., 1996; Welsh and Poe, 1998).[10]. Incentive-compatibility conditions in the elicitation process may also prevent strategic behavior in the sense that the dominant strategy for respondents is to bid truthfully (Wertenbroch and Skiera, 2002). WTP estimated in the laboratory could be subject to hypothetical bias, strategic behaviors, or social desirability bias pushing respondents to overstate their WTP to be more socially acceptable (Lusk and Norwood, 2009; Carlsson et al., 2010; Paulhus, 1991). Leggett et al. (2003) show that WTP values derived from face-to-face interviews could be as much as 23%–29% higher relative to self-administered surveys. We discuss additional moderating variables, which are detailed in section 5.

---

[8] See Barseghyan et al. (2018) for a summary of risk preference models.

[9] Under EUT, a risk-averse agent will buy full insurance if and only if the premium is fair, i.e. equal to expected losses, (Mossin, 1968). For small probability risks, the demand for full insurance at unfair premiums or less than full insurance at fair premiums contradicts the EUT (Schlesinger, 1997).

[10] When asking agents directly about their WTP, they are more likely to dwell on the price or attempt to respond strategically if they believe their responses will affect future pricing (Breidert et al., 2006, Jedidi and Jagpal, 2009). Moreover, simulating real

## 3. The Meta Dataset

### 3.1 Search Strategy and Inclusion Criteria

The first stage in the meta-analysis method consists of selecting the primary studies. To this end, we follow the reporting guidelines and the PRISMA statement outlined in Havranek et al. (2020). We search for empirical studies published in the Web of Science and Google Scholar databases using a combination of the following keywords: "Insurance", "willingness to pay", "low probability", "contingent valuation", "climate risk" and "natural disasters" over the period from 2005 to 2021.[11] We also reviewed the bibliographies of the retrieved papers for more empirical research. Data searches were performed on Harzing's Publish and Perish software to collect primary documents information. To determine which primary studies to include, a list of selection criteria is established. These criteria are necessary to ensure that the final dataset contains studies with a reasonable degree of heterogeneity while still allowing for meaningful comparison. The details of PRISMA steps and results are provided in Figure 1. The identification stage included 15,664 articles identified by the database search (see Appendix B).[12] We remove duplicates and screen articles based on title (1,057 articles). Four main exclusion criteria were applied during the selection phase:

(i) Risk Type: We specifically focused on the nature of risk associated with material losses and damages stemming from idiosyncratic or natural hazards. We exclude studies that primarily dealt with health or life risks, highlighting our dataset on material and financial impacts.

(ii) Loss Probability Threshold: We established a loss probability threshold for inclusion in the study. Specifically, we only included instances where the loss probability was stated was 5% or less.

(iii) Valuation Methodology: We include studies that use contingent valuation-based empirical research for the determination of willingness to pay (WTP). Studies deriving WTP from theoretical modeling or discrete choice experiments were not considered, as our methodology required direct empirical valuation.[13]

(iv) Protection Mechanism: Our focus was on studies that exclusively considered insurance as a protection mechanism against losses. Consequently, we excluded studies that evaluated other forms of financial protection or risk mitigation, such as loans of self-insurance.

For the eligibility step, we included studies that report information about the mean WTP estimates and their corresponding standard errors or variance (standard deviations) estimates.[14] A second important

---

purchasing experiences demands less mental effort than asking respondents to indicate their WTP (Brown et al., 1996). There are also limitations to indirect elicitation approaches that might affect the hypothetical bias. Following Smith et al. (2019) respondents could be more uncertain about their preferences which leads to different responses depending on the question format.

[11] There are two main reasons why we select 2005 as a starting date for the dataset. We try to ensure an adequate representation of the three forms of insurance against low-probability risks i.e. market, micro, and index insurance. Results before 2005 obtained using inclusion/exclusion criteria were very sparse with no studies on micro or index insurance. The second reason pertains to the entry into force of the Kyoto Protocol, which became legally binding on its 128 signatories on February 16, 2005.

[12] We initially conducted separate searches in both Web of Science and Google Scholar. However, after comparing the results from both databases, we found that Google Scholar provided a broader range of articles, including all of the articles identified from Web of Science and additional ones as well. These reported values are related to Google Scholar results.

[13] We acknowledge the importance of discrete choice experiments (DCE) studies, but the decision not to include these studies came about because of data availability and coherence issues. DCE studies indirectly calculate WTP for insurance schemes for specific attributes (e.g., nature of the supplier, deductible amount, coverage level, premium frequency, contract duration, etc.). Because each DCE research has its unique set of attributes, many moderators' observations would be missing, posing a considerable problem for the MRA. The second issue is related to the difficulty of finding reliable information on expected loss for each choice set (with specific attribute levels).

[14] The limitation of obtaining raw data constrained our capacity to compute the WTP and its standard error. Consequently, we resorted to extracting these statistics directly from the included studies, which, resulted in a reduction in the size of our dataset.

criterion is the presence of information to allow the measurement of expected loss, the average historical cost of annual losses or, alternativity the actuarially fair premium. We included peer-reviewed articles published in English or Chinese.[15] To identify additional studies, we reviewed the reference list of retrieved articles. We included 38 studies in our meta-analysis, which led to 74 data points as some studies included multiple observations (see Appendix C for further information).vThe final dataset includes average WTP estimated either from observational or experimental studies. Three separate instances are covered by observational studies. WTP may be elicited using: (1) hypothetical/actual scenarios with information on probability and loss, (2) actual scenarios with no probability information, and (3) no scenario at all with no information (see Appendix D). However, experimental studies encompass two cases. The first estimates the WTP for different levels of probabilities and/or losses. The context description is neutral and the insurance contract is without default risk. For the second type of studies, the WTP is elicited by the manipulation of additional factors (e.g. framing, default probability, etc.) other than probability or losses that are fixed and known throughout the experiment. For these studies, we select only the average WTP elicited from the control group, where the scenario description is neutral and the insurance contract is without default risk.[16],[17]

**Figure 1**: PRISMA flow diagram of the empirical studies included in the meta-analysis

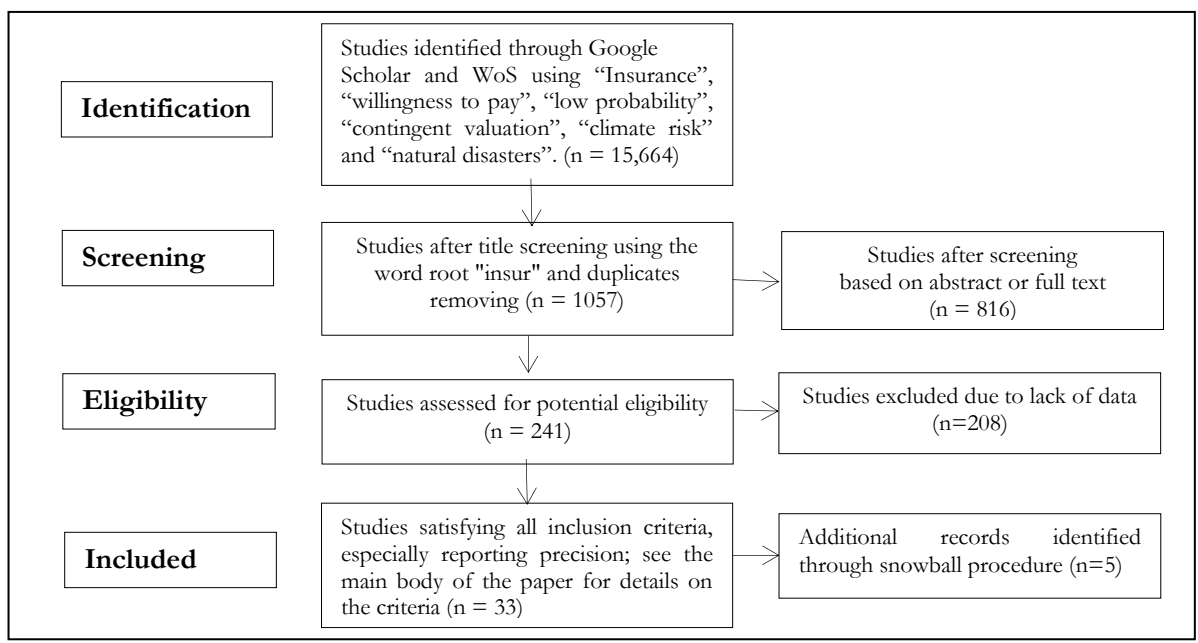

Figure 1 shows the PRISMA diagram of the literature search and selection procedures. The identification stage included 15,664 articles identified by the database search. For the eligibility step, we included studies that report information about the mean WTP estimates and their corresponding standard errors or variance (standard deviations) estimates. We require, as a second criterion, the presence of information to allow the measurement of expected loss, the average historical cost of annual losses or, alternatively the actuarially fair premium. We included peer-reviewed articles published in English or in Chinese. To identify additional studies, we reviewed the reference list of retrieved articles. We included 38 studies in our meta-analysis, which led to 74 data points.

---

[15] As most research on developing countries is done in China, we include articles published in Chinese (see Appendix D).

[16] Experimental findings reveal that individuals show a marked response to non-performance risks (like insurer default or claim issues), leading to a roughly 20% decrease in their WTP, diverging from EUT predictions as noted by Wakker et al. (1997).

[17] SM and NHL carried out the searching, reviewing/screening, and coding of the literature tasks. To reduce error, we worked independently to search for the relevant literature based on different combinations of keywords as specified in the core text. At the end of this phase, results were compared, and duplicates were removed. For the second step of the process, we also worked independently on reading, filtering, and choosing articles. Based on inclusion/exclusion criteria, we note a satisfactory consensus level about the final dataset near 97%. Related to coding relevant variables, SM achieved at first this task by reporting in an Excel file numerical information related to different dependent and moderating variables. NHL performed a double-check of all the coded variables by re-conducting the work from the beginning. It should also be noted that minor discrepancies were discussed by the authors and resolved by consensus or by recourse to the third author (SMA).

**3.2 Effect Size**

As a natural measure of low-probability insurance elicited value, the average willingness to pay denoted by $\overline{WTP}$ would be the most obvious. Expressed in real or experimental monetary units across studies, this metric needs to be normalized. A first concern of such conversion is that the final metric will poorly represent the potential underinsurance dynamics against low probability risks. Furthermore, it may also introduce additional noise into the WTP estimates due to unobserved heterogeneity related, for instance, to variation in risk characteristics, relative cost of insurance or cultural risk and insurance perception, etc. As this kind of heterogeneity would be under-captured by the meta-regression analysis, baseline conversion would not be optimal.[18] To get more comparable outcomes and overcome these particular issues, we define the average willingness to pay per monetary unit of expected loss. We then refer to this metric as relative willingness to pay (RWTP), defined as[19]:

$$RWTP_i = \frac{\overline{WTP_i}}{EL_i} \qquad (2)$$

where $EL_i$ denotes the expected loss for the study ($i$). We assume that expected loss can be either (1) estimated from historical average losses, (2) calculated from known loss distributions or (3) measured from publicly available information on actuarially fair premium. The expected loss information is manually collected from included studies (see Appendix C for details). Adjusting the average willingness to pay with expected loss, as an alternative to statistical rescaling, produces a unitless index of insurance value with one as a reference level. If $\overline{WTP}$ substantially marks up expected loss levels, this may suggest a high value placed on insurance, and vice versa. Figure 2 illustrates the zero-truncated distribution of RWTP across all studies included in our meta-analysis. We note that the distribution is bimodal right-skewed, with 70.6 % of observations less than one, and 13.8% more than two.

**Figure 2**: Histogram of the relative willingness to pay (RWTP) for LPHI insurance

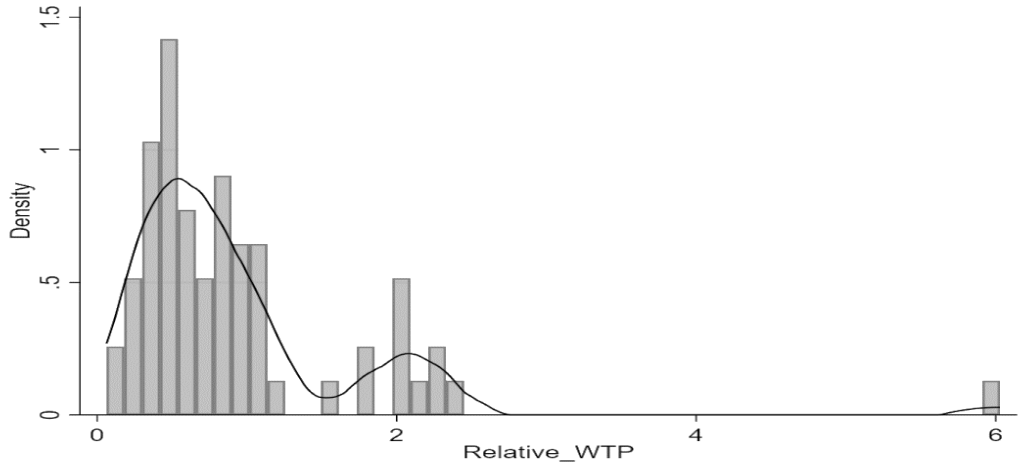

Figure 2 shows the histogram and the corresponding fitted distribution of the relative WTP defined in equation (2) as the mean WTP divided by the expected loss.

---

[18] Standardizing average WTP by its standard deviation eliminates the original units' issue by expressing estimates in relative terms. However, standard deviations are highly dependent on the variable scale. In addition, studies using different experimental designs will have different standard deviation values, which will reduce comparability (Morris and DeShon, 2002). Standard deviations may also be more subject to publication bias in that studies with large standard errors produce estimates with large confidence intervals and would be more difficult to publish.

[19] The RWTP bears some similarity with the response ratio (RR) as defined by Schmidt and Bijmolt (2020). However, it is important to be cautious when comparing these two measures due to their inherent differences.

The standard error of the metric varies considerably in our dataset, raising concerns about outliers that could distort the validity and robustness of the meta-analysis conclusions (Viechtbauer and Cheung, 2010). To alleviate this problem, we trim the RWTP and standard error at the top of the 5% level, resulting in a final dataset of 65 observations.[20] Table 1 presents the mean RWTP values for different sub-groups.[21] The first column reports the unweighted means, while the second reports the weighted means. The meta-analytic mean of RWTP weighted by the inverse of the number of reported observations is about 0.875. At this level, the overall average value should be interpreted with caution because of potential publication bias and heterogeneity examined in the next section. It is interesting to note that experimental-based studies report higher values for RWTP than observational survey studies. Similarly, studies related to idiosyncratic risks appear to report higher values for RWTP than correlated risks. The difference between China, on the one hand, and Germany and the Netherlands, on the other, is significant. Finally, laboratory-elicited RWTP shows the highest values.

**Table 1**. Full sample and subsample RWTP
This table presents the mean RWTP values for the full sample and different sub-groups. The first column reports the unweighted mean, while the second reports the weighted mean.

|  | Unweighted mean | Weighted mean | Observations |
|---|---|---|---|
| Full sample | 0.944 | 0.875 | 65 |
| Subsample observations |  |  |  |
| Survey | 0.613 | 0.641 | 40 |
| Experiment | 1.473 | 1.455 | 25 |
| Idiosyncratic risk | 1.534 | 1.672 | 14 |
| Correlated risk | 0.782 | 0.707 | 51 |
| China | 0.433 | 0.447 | 21 |
| Germany | 1.463 | 1.592 | 13 |
| Netherlands | 1.358 | 1.022 | 12 |
| Laboratory | 1.787 | 1.783 | 8 |

**Note**: In weighted means, RWTP values are weighted by the inverse of the number of estimates per study. We trim the RWTP and the standard error at the top of the 5% level. The final dataset includes 65 observations.

In our analysis of publication bias and meta-regression analysis, we use logarithmic transformation for RWTP.[22] We define the effect size by the standardized willingness to pay (Henceforth SWPT) as follows:[23]

$$SWTP_i = \ln\left(\frac{\overline{WTP_i}}{EL_i}\right) \qquad (3)$$

The standard error of SWTP:[24]

$$SE(RR_i) = \sqrt{\frac{SD_i^2}{n_i \overline{WTP_i}^2}} \qquad (4)$$

where $SD_i$ denotes the standard deviation of the WTP for study ($i$) and $n_i$ is the sample size.

---

[20] Winsorization is an alternative method to apply in the meta-analysis (Lipsey and Wilson, 2001). It substitutes the extreme values with the highest values in given percentiles.

[21] Appendix D reports the relative frequencies of the included studies, using bar diagrams, according to different subgroups (continent, year, country, coverage type, elicitation method, insurance type, risk type, and number of studies).

[22] Logarithm transformation linearizes the RWTP metric so that deviations in the numerator and denominator have the same impact (Hedges et al., 1999). Additionally, the moderating variables coefficients in the meta-regression would be easier to interpret. Third, the distribution of the logarithm of response ratios is approximately normally distributed (Hedges et al., 1999). The absence of such normalization has a minor impact on the estimates of meta-regression coefficients discussed in section 5.

[23] The sample mean, as a measure of central tendency, does not quantify a causal relationship between two variables of interest, and thus there is no "effect", Borenstein et al. (2011). For convenience, we will use the terms "metric" and "effect size" interchangeably to represent mean WTP adjusted by the expected loss for the insured risk.

[24] See Appendix A for the derivation of equations 3 et 4.

## 4. Publication Bias

Publication bias is a perennial concern that may distort the estimation of the average overall effect and the conclusions drawn. The studies included in the current meta-analysis are observational and non-comparative, so the interpretation of outcomes would not be contingent on the null hypothesis significance test (Maulik et al., 2011). In practice, studies reporting low WTP are equally likely to be published as those with very high WTP, provided they meet rigorous standards. It is not uncommon, that for willingness to pay (WTP) elicitation to lack rigor or to take insufficient account of hypothetical biases or strategic behaviors. These methodological shortcomings could reduce a study's chances of publication, even if realistic WTP values are obtained from a reasonable sample size. Similarly, studies with lower WTP variance may be less likely to be published because of the greater difficulty in accurately predicting WTP.

A starting point to get some insight into the presence of publication bias is a visual inspection of the funnel plot, which presents SWPT on the horizontal axis and the precision of the estimates (1/SE) on the vertical axis. If the distribution of standard errors is symmetrically distributed around the mean line, there is no publication bias. Figure 3 shows the funnel plot for the assessment of publication bias. The shape of the funnel plot did not reveal any evidence of apparent visual asymmetry. A more formal and accurate way to detect publication bias is the "Funnel Asymmetry Test"-"Precision Effect Test" (FAT-PET) proposed by Stanley (2008). This test assumes that publication selection induces a correlation between the estimated effect size and their standard errors. The FAT-PET is implemented by testing the slope of the regression of SWTP on its standard error:

$$SWTP_i = \alpha_0 + \alpha_1 SE(SWTP_i) + \varepsilon_i \qquad (5)$$

where SWTP is the $i^{th}$ standardized WTP estimated in study $s$ and $SE(SWTP)$ is the corresponding standard error, $\alpha_0$ is the true effect after correcting for publication effect and $\alpha_1$ is a measure of the importance of publication bias. Testing for $\alpha_0 = 0$ is a precision effect test (PET) for a genuine empirical effect net of publication bias, whilst testing for $\alpha_1 = 0$ is the funnel asymmetry test (FAT) for publication selection. The statistical significance of the estimate of $\alpha_1$ is an indicator of the presence of publication bias. Since the empirical studies in our dataset use different data collection methods and sample sizes, $\varepsilon_i$ are likely to be heteroscedastic. Equation 5 is thus estimated using the weighted least square (WLS) method[25] using Fixed effects (FE) and Random effects (RE) models.

**Figure 3.** Funnel Plot of the SWTP

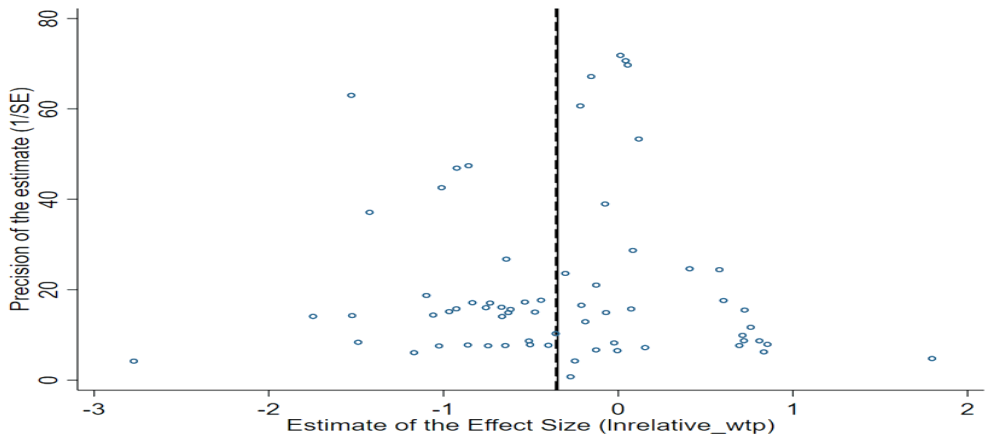

Figure 3 shows the funnel plot for the assessment of publication bias. The horizontal axis represents the SWTP values. The vertical axis represents the inverse of the standard errors.

---

[25] We further discuss the heteroscedasticity issue in section 5 when we produce heterogeneity analysis.

The fixed-effects model assumes that the effect sizes of the studies are deterministic and different. The random effects model assumes, however, that true effects can differ across studies so that the variation in estimated effects is composed of two parts: heterogeneity (between studies) in the true effect and sampling error. The weight is thus $1/(SE_i^2 + \tau^2)$ where $\tau^2$ measures the variance of the true effect in the population (often referred to as the amount of 'heterogeneity' in the true outcomes).[26]

Columns 1 to 3 of Table 2 present the results of three specifications based on equation (5): simple OLS, WLS-RE and WLS-FE. To accommodate within-study correlation of estimates for each specification, we report cluster-robust standard errors with clustering by study. Moreover, two weighting schemes are used for each specification: equal weights for each estimate (weight 1) and equal weights for each study (weight 2). The second weighting scheme allows the consideration of multiple effect-size estimates reported by primary studies.[27] For all specifications, we do not reject the null hypothesis for $\alpha_1$, which indicates the absence of publication bias: ($\alpha_1 = 0$ at the 10% significance level). Stanley (2008) argues that the publication bias-corrected estimates of the mean true effect ($\alpha_0$ in Equation (1)) may be biased downward when the null hypothesis is rejected. While the null hypothesis is not rejected, we follow the procedure proposed by Stanley and Doucouliagos (2014) that consists of replacing the standard error with its squared term (quadratic specification), i.e., the variance. The meta-regression is called in this case the Precision Effect Estimate with Standard Error (PEESE). Columns 4 to 6 of Table 2 display the PEESE results. We find the same results as columns 1 to 3, that is there is no publication bias in all specifications (OLS and WLS-RE in the two-weighting scheme).[28]

**Table 2:** FAT-PET and PEESE of publication bias (WTP)

Table 2 presents the results of the FAT-PET and PEESE tests of publication bias. We use three specifications of the model formulated in equation (5): simple OLS, WLS-RE, and WLS-FE.

|  | OLS | WLS-RE | WLS-FE | OLS | WLS-RE | WLS-FE |
|---|---|---|---|---|---|---|
|  | FAT-PET | FAT-PET | FAT-PET | PEESE | PEESE | PEESE |
|  | (1) | (2) | (3) | (4) | (5) | (6) |
| Weight 1: Equal weight to each estimate | | | | | | |
| $\alpha_1$ (**Pu**b. bias) | 0.1418 | 0.3389 | -0.8717 | 0.0438 | 0.056 | 0.246 |
|  | (0.316) | (1.029) | (4.094) | (0.0652) | (0.157) | (5.265) |
| $\alpha_0$ (**Precision**) | **-0.36**** | **-0.3768**** | **-0.3319** | **-0.3482**** | **-0.3487**** | **-0.3558*** |
|  | (0.1342) | (0.1288) | (0.2951) | (0.1409) | (0.14) | (0.208) |
| Observations | 65 | 65 | 65 | 65 | 65 | 65 |
| R-squared | 0.001 | 0.176 | 0.127 | 0.0002 | 0.175 | 0.275 |
| Weight 2: Equal weight to each study | | | | | | |
| $\alpha_1$ (**Pu**b. bias) | 0.09 | 0.104 | 4.215 | 0.0593 | 0.046 | 2.7269 |
|  | (0.1582) | (0.46) | (5.383) | (0.0699) | (0.0908) | (4.10) |
| $\alpha_0$ (**Precision**) | **-0.396**** | **-0.396**** | **-0.6713*** | **-0.3892**** | **-0.3885**** | **-0.544**** |
|  | (0.1332) | (0.1357) | (0.367) | (0.1269) | (0.1266) | (0.227) |
| Observations | 65 | 65 | 65 | 65 | 65 | 65 |
| R-squared | 0.002 | 0.241 | 0.280 | 0.001 | 0.240 | 0.404 |

**Notes:** Robust standard errors clustered at the study level are shown in parentheses. OLS = ordinary least squares, WLS-FE = weighted least square fixed effects; WLS-RE = weighted least square-random effects. *, **, and *** denote statistical significance at 10%, 1%, and .1%, respectively.

---

[26] We use the restricted maximum–likelihood (REML) estimator and the Knapp-Hartung standard-error adjustment for the estimation of $\tau^2$. Our results are robust to other estimators (such as the Hedges estimator, the Šidák–Jonkman estimator, and the DerSimonian–Laird estimator).

[27] To mitigate the domination effect of studies with a large number of estimates, we estimate equation (5) with frequency weights, specified as the inverse of the number of estimates reported in each study.

[28] We conduct an additional test to measure the bias-adjusted true effect ($\alpha_0$), the weighted average of adequately powered (WAAP) estimator by Ioannidis et al. (2017). As expected, we find no evidence for publication bias (results available upon request).

## 5. Heterogeneity Analysis

### 5.1 Variables Description

The average RWTP varies considerably across studies, as shown in Figure 2. The null hypothesis of Cochran's Q test reported in Appendix E indicates a large level of heterogeneity between studies ($I^2 > 75\%$). To deal with heterogeneity and to identify the most effective factors that would explain differences between RWTP, we define several moderating variables (binary, multinomial, and numeric) as covariates in the meta-regression. As a second objective, we define two synthetic study profiles that simulate an average RWTP using all estimates, but overweighting those that are better identified.

We separate the moderators into the following categories: WTP elicitation design, risk specificities, exposed assets, insurance features, sample respondents' characteristics, spatial-temporal variations, and publication characteristics. Table 3 presents the definition and summary statistics of all variables included for heterogeneity analysis. In the first category, we consider moderators that focus on the survey design and measurement characteristics of WTP. As reported in Table 2, this first category represents an important source of heterogeneity. We distinguish observational surveys 62% of our dataset from controlled experiments 38%, which breaks down to 25% for online experiments, 12% for laboratory experiments, and 1% for field experiments. For some studies, there are several scenarios where authors use between-subjects or within-subjects designs.

For the WTP measurement methods, we describe the prevalence of compatible incentive mechanisms by a binary variable with an average value of 18%. We also consider the fact that WTP is measured using a price generation approach (e.g., an open-ended question) as opposed to a price selection approach (simple or double dichotomy method). Hypothetical bias mitigation correction is modeled by a binary variable that indicates whether researchers employ bias mitigation strategies (e.g. cheap talk, consequential script, follow-up question, etc.). The participation fees binary variable indicates whether participants received monetary compensation for their participation in the study.

We encoded the variability of the risks by considering the difference between idiosyncratic risks (21.5%) and correlated risks (78.5%). For the second category, we specify different subclasses (flood 61%, various climatic risks 9%, and earthquake 1%). For the risk characteristics, we define a first numeric variable equal to the descriptive probability provided and a second binary variable that describes the studies in which the implicit probability of loss is below the 5% threshold. When the probability information is not provided (68% of the cases), we estimate it from the fair premium or the average loss. Regarding the third category, we note that assets exposed to small probability risks are disparate. We define two main classes of assets: crops and property (house and contents), which account for 40% and 32% of our sample, respectively. Regarding insurance contracts, 60% (40%) are indemnity-based (index-based), whereas 29% have a subsidized premium. Sample and data characteristics include a set of dummy variables to indicate whether the estimates are related to the entire population or from targeting populations at risk (51% of the dataset). We also code two binary moderators for studies that distinguish between "protest" and "true zero" WTP. We create a set of variables related to the main countries of study, which are China, Germany, and the Netherlands with the presence frequencies of 40%, 20%; and 18% respectively, as depicted in Figure 4. The year of data collection is included, with the distribution of this data depicted in Figure 5. We introduce two variables related to average age and average annual income converted to US dollars using the corresponding exchange rates.

The last category of moderators contains publication characteristics and relies on four variables. The first one is the number of citations to account for study quality. A second variable indicates whether the study was published in an international academic journal recognized by the French National Research Center (CNRS). We also denote by a binary variable the studies with low citations (less or equal to one). Finally, we perform a diagnostic test for multi-collinearity on all variables. The values of the variance-inflation (VIF) factors for all variables are lower than 9, with an average VIF of less than 5 (see Appendix F).

**Table 3. Description and summary statistics of variables**

This table presents the main metric and the moderator variables and describes their construction. The table also presents summary statistics, including unweighted mean, standard deviation, and weighted mean. The variables are aggregated into different categories (WTP measurement design, risk types & characteristics, exposed assets and insurance characteristics, sample characteristics, study regions and year, and publication characteristics).

| Variable | Description | Mean | Std. deviation | Weighted Mean |
|---|---|---|---|---|
| SWTP | The logarithm of average WTP divided by the expected loss | -0.346 | 0.763 | -0.385 |
| Standard error | Standard error of SWTP | 0.101 | 0.165 | 0.112 |
| *WTP measurement design* | | | | |
| Observational survey | =1 if the estimate is from observational survey data, 0 otherwise | 0.615 | 0.50 | 0.713 |
| Experience | =1 if the estimate is from experience data, 0 otherwise | 0.384 | 0.49 | 0.286 |
| Lab | =1 if the estimate is from laboratory experience, 0 otherwise | 0.123 | 0.33 | 0.137 |
| Field | =1 if the estimate is from field experience, 0 otherwise | 0.015 | 0.124 | .0274 |
| Online | =1 if the estimate is elicited online, 0 otherwise | 0.246 | 0.434 | 0.122 |
| Within design | =1 if the estimate is from within subjects' design, 0 otherwise | 0.353 | 0.481 | 0.181 |
| Incentive compatible | =1 if the elicitation is incentive compatible, 0 otherwise | 0.184 | 0.391 | 0.126 |
| HB mitigation | =1 if there is hypothetical bias mitigation, 0 otherwise | 0.169 | 0.388 | 0.132 |
| Elicitation method | =0 if WTP is generated through direct elicitation methods<br>=1 if WTP is generated through hybrid elicitation methods<br>=2 if WTP is generated through indirect elicitation methods | 0.415 | 0.496 | 0.521 |
| Participant fee/show-up | =1 if participants received participation fee or show-up, 0 otherwise | 0.23 | 0.424 | 0.209 |
| *Risk types and characteristics* | | | | |
| Idiosyncratic risk | =1 if the study considers no correlated losses, 0 otherwise | 0.221 | 0.414 | 0.173 |
| Climate risk | =1 if the study examines coverage against climate risk, 0 otherwise | 0.092 | 0.291 | 0.137 |
| Earthquake risk | =1 if the study examines coverage against Earthquake risk, 0 otherwise | 0.015 | 0.124 | 0.027 |
| Flood risk | =1 if the study examines coverage against Flood risk, 0 otherwise | 0.615 | 0.49 | 0.593 |
| Implicit probability | =1 if the probability of loss is not explicitly provided in the study, 0 otherwise | 0.323 | 0.471 | 0.291 |
| Probability level | Value of the provided probability of loss | 0.006 | 0.015 | 0.004 |

| Variable | Description | Mean | Std. dev. | Weighted Mean |
|---|---|---|---|---|
| **Exposure assets and insurance characteristics** | | | | |
| House | =1 if the exposed asset is a property (house/contents), 0 otherwise | 0.323 | 0.471 | 0.232 |
| Crop | =1 if the exposed asset is a crop, 0 otherwise | 0.4 | 0.493 | 0.438 |
| Indemnity insurance | =1 if the study considers indemnity insurance, 0 otherwise | 0.6 | 0.493 | 0.561 |
| Presence of subsidy | =1 if there is an insurance premium subsidy, 0 otherwise | 0.292 | 0.458 | 0.246 |
| **Sample characteristics** | | | | |
| Subject pool | =1 if the WTP is estimated from the general population, 0 otherwise | 0.293 | 0.458 | 0.256 |
| Random sample | =1 if the WTP is estimated from a random sample, 0 otherwise | 0.923 | 0.268 | 0.89 |
| Sample size | Number of observations of the study | 348.18 | 369.4 | 382.82 |
| Protest zeros WTP | =1 if the study accounts for protest zeros WTP, 0 otherwise | 0.415 | 0.496 | 0.401 |
| **Regions and study year** | | | | |
| China | =1 if the study is realized in China, 0 otherwise | 0.323 | 0.471 | 0.301 |
| Germany | =1 if the study is realized in Germany, 0 otherwise | 0.200 | 0.400 | 0.132 |
| Netherlands | =1 if the study is realized in the Netherlands, 0 otherwise | 0.180 | 0.391 | 0.113 |
| Year | The year the study was conducted | 2011.7 | 4.6 | 2012.67 |
| **Control variables** | | | | |
| Annual income | The logarithm of sample annual income in U.S. dollars (inflation-adjusted) | 0.311 | 0.54 | 0.357 |
| Average age | Sample average age in years | 43.72 | 8.357 | 44.69 |
| **Publication characteristics** | | | | |
| Article type | =1 if the study is published, 0 if it is a working paper | 0.953 | 0.211 | 0.972 |
| Top-ranked academic journal | =1 if the study is published in an acknowledged academic journal, 0 otherwise | 0.323 | 0.471 | 0.324 |
| Low number of citations | =1 if the average number of citations per year is less than one, 0 otherwise | 0.261 | 0.442 | 0.246 |

**Notes:** The third column corresponds to the mean weighted by the inverse of the number of estimates per study.

**Figure 4: Distribution of studies by country**

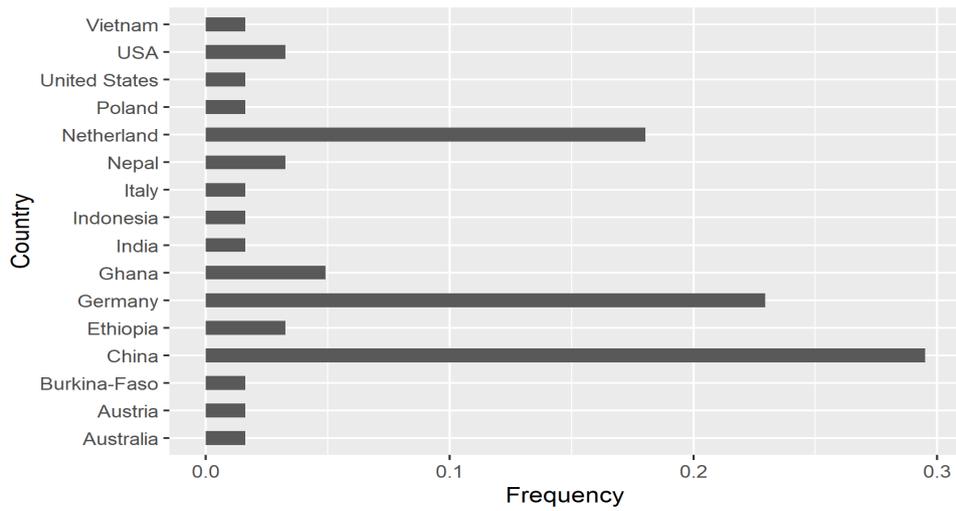

This figure shows the relative frequencies of in included studies (n=65) by country.

**Figure 5: Distribution of studies per year**

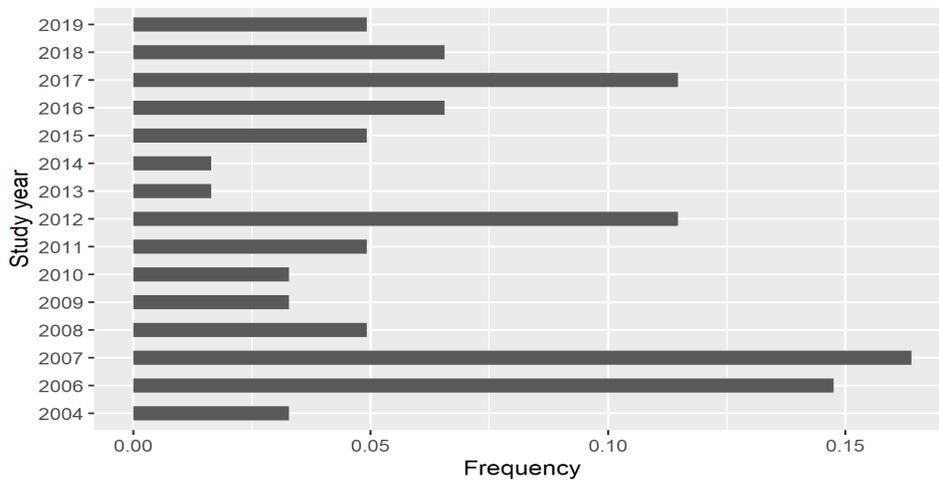

This figure shows the relative frequencies of in included studies (n=65) by year of data collection.

**5.2 Meta-Regression Model**

We investigate potential sources of heterogeneity by completing the model provided in equation (5) with additional study-level characteristics. We intend to estimate the "True" SWTP level after accounting for the potential effect of moderating variables. The baseline meta-regression model is then formulated as follows:

$$SWTP_{i,s} = \alpha_0 + \alpha_1 SE(SWTP_{i,s}) + X_{i,s}\beta + \varepsilon_{i,s} \quad (6)$$

With $SWTP_i$ the logarithm of the mean WTP divided by expected loss, $X$ a vector of variables (moderating variables) to capture study-specific characteristics associated with the estimate $s$ from study $i$, $\beta$ a vector of coefficients and $\varepsilon_i$ the sampling error of the regression. The intercept term of the meta-regression, $\alpha_0$, measures the true level of SWTP after controlling for publication bias and heterogeneity, Xue et al. (2021). A statistically insignificant intercept means that the observed estimates are driven mainly by the characteristics of the primary studies. Conversely, a statistically significant intercept may suggest an intrinsic perceived value of insurance irrespective of the potential effect of moderating variables.

Unlike conventional econometric models, we cannot assume that the estimation errors of the meta-regression model are independent and identically distributed. First, dependence is likely to arise, especially when there are multiple estimates from a unique study (Stanley and Doucouliagos, 2014). In such a case, this study's results might dominate the overall effect. Second, heteroscedasticity, i.e. non-constant variance of SWTP estimates, could also be present due to primary studies using different sample sizes, sample randomness, and sampling method (Nelson and Kennedy, 2009). Therefore, estimating a meta-regression with the OLS method might lead to inconsistent estimates, though unbiased. For these reasons, we estimate meta-regression with the Weighted-least squares (WLS) method. We perform meta-regression using three estimators: (1) a cluster-robust ordinary least squares (OLS) estimator; (2) a cluster-robust random effects model (Unweighted RE) (3) a weighted random effects model by the inverse of the standard error (Weighted RE).[29] The choice between the two models depends on how the individual studies are collected. If the effect size is identical across the studies a fixed-effects model can be used (Hedges & Vevea, 1998). When there is significant heterogeneity among the studies included in the meta-analysis a random-effects model is preferred. The high heterogeneity of our dataset in terms of research design, time of publication, data collection, and national context lead us to prefer a random-effects model (see the Q statistic Appendix E). For robustness reasons, we run analyses using other models.[30]

Meta-regression faces the so-called "model uncertainty" problem, which implies that the true model cannot be identified in advance, Brada et al. (2021) and Kocenda and Iwasaki (2022). Having the wrong variables in the regression model leads to misspecification bias and invalid inference. To address this problem, we estimated our models using the Bayesian model averaging (BMA). The objective of this method is to define the best possible approximation of the distribution of regression parameters. BMA analysis provides three basic statistics for each parameter: the posterior mean, the posterior variance, and the posterior inclusion probability. The likelihood of each model is reflected by the model's posterior probabilities. The posterior means are then calculated as the estimated coefficients weighted across all models by their posterior model probability. We follow Jeffreys (1961) to interpret the posterior inclusion probabilities (PIPs) of BMA means, who characterizes evidence of an effect as "weak" for a PIP between 0.5 and 0.75, "substantial" for a PIP between 0.75 and 0.95, "strong" for a PIP between 0.95 and 0.99, and "decisive" for a PIP above 0.99.[31]

---

[29] The random-effect (RE) model is more appropriate in the presence of high heterogeneity. RE model weights correspond to $1/(\tau^2+SE_i^2)$. When heterogeneity is high, $SE_i^2$ would be negligible compared to $\tau^2$ so that all data points would have the same weight≈$1/\tau^2$, which can be problematic in the presence of publication bias. We weight the dataset by $1/SE_i^2$ before using the RE model.

[30] We estimate the random/fixed effects model using the R package "metaphor" (rma, REML). The Bayesian Model Average is estimated with R package "bms" and the unweighted OLS with R package "Robustbase" (lmrob).

## 5.3 Results

Table 4 reports the meta-regression analysis (MRA) obtained from different estimation methods after checking for multi-collinearity. Consistent with the previous results presented in Table 2, coefficients associated with publication bias are not statistically significant in all estimated models. A first view of results is provided in Figure 6 which depicts a visual representation of the BMA analysis. We note that 14 moderating variables are relevant to explain the observed heterogeneity across studies with a PIP higher than 0.5.[32] From Table 4, estimated coefficients vary relatively little across the different models showing very small differences between the weighted (1 & 2) and unweighted (3) estimation methods.[33] Following a series of preliminary analyses, a total of 23 moderating variables have been retained in MRA to mitigate concerns regarding overfitting.[34]

Following the MRA, we focus on moderators for which we have the strongest effect on the SWTP i.e. the highest posterior inclusion probability in BMA analysis. The absence of probability of loss information provision is linked to lower SWTP. When loss probability is not provided, individuals tend to form their estimations through sampling (Barron and Erev 2003; Hertwig et al., 2004; Weber et al. 2004, Bakkensen and Barrage, 2021). Due to small experienced samples, losses will be seldom, implying an underweighted chance of losses and a consequently low insurance demand. Krawczyk et al. (2017) highlight a persistent underestimation of small probability risks, even when subjects learn about the risk over time. On the other hand, when probabilities are provided, we find that a small decrease can have a significant positive effect on SWTP. All else being equal, a 1% downward variation of the provided probability tends to increase SWTP by 0.26, suggesting that the decay rate of WTP, for a small probability decrease, is less than the expected loss. This result seems to contradict the probability neglect idea, suggesting that provided small probabilities are above the threshold of concern. A similar result was found for stated WTP for insurance, with very small provided probabilities, in Schade et al., (2012). This result is remarkably consistent with the description-experience decision gap, where behavioral implications vary depending on whether uncertain choices are made from experience or description (Hertwig et al., 2009). For description-based prospects, people seem to be risk averse for gains and risk seeking for losses. However, for an experience-based setting, they become risk-seeking for gains and risk-averse for losses (Kudryavtsev and Pavlodsky, 2012). Decisions from description (DFD), where an explicit and precise description of the loss probability distribution is provided to agents, are more subject to small probability distortion.

---

[31] BMA requires explicit priors on the model (model prior) and regression coefficients (g-prior). As suggested by Eicher et al. (2011), we use the uniform model prior and the unitary g-prior information.

[32] The vertical axis lists all our moderating variables sorted by the posterior inclusion probability (PIP) in descending order and the horizontal axis refers to the posterior model probability (PMP) of each model sorted in ascending order. The blue and red colors indicate positive and negative signs of moderatos, respectively. The blank cells suggest that the parameters associated with these variables are not significantly different from zero for most models.

[33] We estimate the random effects model using the R package "metaphor" (rma, REML) without and with weighting by ($1/SE_i^2$). The Bayesian Model Average is estimated with R package "bms" and the unweighted OLS with R package "Robustbase" (lmrob).

[34] The debate related to the number of studies per covariate remains unresolved between the traditional rules of thumb used to minimize the risk of overfitting and the new findings. We draw on the work of Austin et al. and Steyerberg (2015), who suggest that the number of observations required per covariate may be lower than often assumed. They suggest that as few as two studies per variable might suffice for a reasonable estimation of regression coefficients.

*Figure* **6**: Visual Representation of BMA

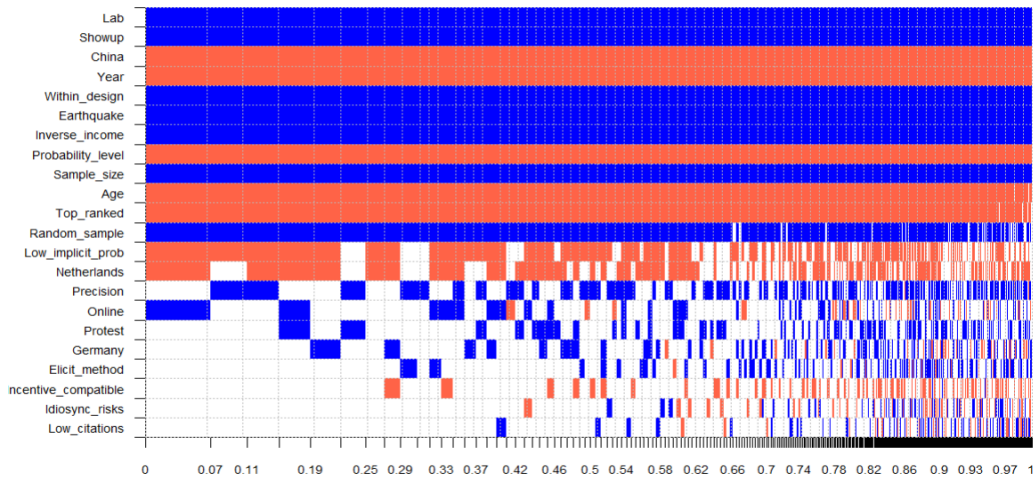

This figure shows the graphical result of meta-regression using the Bayesian model averaging. Numerical results are reported in Table 4. All variables are described in Table 3. On the vertical axis, the explanatory variables are ranked according to their posterior inclusion probabilities from the highest at the top to the lowest at the bottom. The horizontal axis shows the values of cumulative posterior model probability. Blue color (darker in grayscale) = the estimated parameter of a corresponding explanatory variable is positive. Red color (lighter in grayscale) = the estimated parameter of a corresponding explanatory variable is negative. No color = the corresponding explanatory variable is not included in the model.

This result also seems consistent with the inverse S-shaped probability weighting heuristic and, in particular, with the local condition that extremely low probabilities are more overweighed relative to their base value than small probabilities (Jaspersen et al., 2023).[35]

While our results provide no direct support for hypothetical bias, additional factors related to data collection design seem to impact the SWTP. For instance, the laboratory-based elicitation method shows a remarkably positive effect on the outcome variable, in contrast to online experiments and surveys. Laboratory-based methods often involve a hypothetical purchase decision of hypothetical/real insurance products without monetary consequences. Estimated WTP could be subject to hypothetical bias, strategic behaviors, or social desirability bias that may push respondents to overstate their WTP to be more socially acceptable (Lusk and Norwood, 2009; Carlsson et al., 2010; Paulhus, 1991). MRA does not provide evidence on the influence of incentive-based elicitation settings, and the relative effects of direct versus indirect elicitation methods. Such variables are generally considered proxies for hypothetical biases and strategic behaviors. These results are in line with Kesternich et al. (2013) who find no difference between market data and hypothetical choice experiments when estimating Medicare insurance demand. Similarly, Cole et al. (2020) provide no support for hypothetical bias when estimating demand with different elicitation mechanisms.[36] Other moderators such as participation (show-up) fees and multiple WTP estimates based on within-subject design report significant positive effects on SWTP. Within-subject design can be more prone to the potential correlation of treatment effects and to other systematic behavioral patterns such as learning, subject fatigue, wealth effects, etc., (Landry, 2017).

Our results do not show evidence that correlated risks are related to low WTP as opposed to the claim of Friedl et al. (2014) who show that social comparisons make insurance less attractive when risks are correlated because of subjects' aversion to unequal payoffs.[37] To further investigate the effect of correlated risks, we analyze the potential effect of various sub-categories of correlated risks observed in our dataset such as climate, flood, snow and earthquake. It turns out the SWTP for earthquakes is significantly positive. This result should be further explored in future studies.

---

[35] This outcome is particularly reconcilable with the Jaspersen et al. (2023) model and their new local condition of the probability weighting function, the decreasing relative overweight (DRO).

[36] All these results are in line with the earlier Loomis (2011) findings where hypothetical bias is less severe for WTP elicited for private goods. However, Robinson and Botzen (2018) find some evidence for hypothetical bias in their results related to the insurance context.

[37] One possible explanation of this stylized fact may lie in the difficulty of indemnifying concomitant and correlated losses, resulting in fat tails loss distribution which might threaten the solvency of insurers and their ability to fulfill their contractual commitments to policyholders (Biener et al., 2019).

Average sample incomes are negatively associated with SWTP supporting the view of insurance as an inferior good. This result may suggest that wealthier individuals would face lower costs to self-insure, an activity that substitutes for market insurance (Ehrlich and Becker, 1972). A theoretical justification for this finding is also related to the decreasing absolute risk aversion for wealth leading to consider insurance as an inferior good. Age is negatively related to the outcome variable in the sense that samples with a high average age exhibited the lowest SWTP. This result confirms the finding of Browne et al. (2015) related to revealed preference analysis for purchasing flood versus bicycle theft insurance. Their results show that demand for both types of coverage decreases with age.[38]

Our results also show that the SWTP tends to decrease with time, a result that echoes persistent underinsurance behavior reported in the literature. This finding might also be indicative of a gradual decrease in attention to low-probability risks or disinterest for market insurance. It is plausible to think that people seem to steadily accept this perception and the consequences of LPHI risks such as climate change. Such conclusions would point in the direction of two antagonist behaviors, the search for alternative cost-effective solutions in which they would have a more proactive role or the acceptance of some degree of vulnerability (Leiserowitz et al., 2019 and Wagner, 2022).[39] The MRA results also show that the relative WTP are geographically dependent with a first notable difference between Germany and the Netherlands. Seifert et al. (2013) compared insurance demand in Europe using data from these two countries and found that WTP is higher in Germany. The charity hazard resulting from the disparity in disaster insurance systems is a possible explanation for this gap (Browne and Hoyt, 2000). Post-disaster public funding as in the Netherlands may encourage individuals to expect to receive contributions from public relief money in the event of a major disaster (Yan and Faure, 2021).

Second, we find that the relative WTP is smaller in China than in Europe, mainly in Germany and the Netherlands. This finding is consistent with several empirical results of studies conducted in developing countries. After controlling for the income effect, one plausible explanation is that for a collectivity-oriented society as in China, post-loss financing may rely on informal family and community solidarity more than on formal market insurance products. Therefore, private insurance may play a smaller role as a risk management mechanism. To reinforce this result, we conduct a complementary analysis using national culture as an additional determinant of perceived costs and benefits of insurance.

To capture national culture, we use Hofstede's cultural factors.[40] Holding other factors constant, we find that relative WTP is positively (negatively) related to "Uncertainty avoidance" ("Power distance"). In countries with high "Power distance", people tend to accept inequalities more easily, reflecting a high degree of centralization of authority. This may also imply an increased reliance on an autocratic social order to manage the post-loss consequences in case of extreme losses, thus reducing the demand for insurance. Countries with high uncertainty avoidance, such as Germany and the Netherlands, are well organized to manage risks through developed insurance markets. In these countries, the willingness to pay to reduce risk through insurance is more significant.[41] Chui and Kwok (2008) and Park and Lemaire (2012) report similar results related to the demand for life and non-life insurance, respectively.

We perform an additional analysis using general-to-specific stepwise regression for further robustness checks. As shown in Table 5, after accounting progressively for key moderators derived from the BMA, remain largely significant.[42] Note that our results remained consistent even when outliers were not excluded, as shown in Appendix L.

---

[38] For stated preference surveys, age may have a significant influence on protest responses. Mental abilities decline with age and cognitive effort needed for making decisions in hypothetical scenarios may lead to more protest responses. Accordingly, younger individuals may be more likely to accept hypothetical scenarios than older individuals.

[39] Further research is needed to figure out if the public mechanisms that are currently in place to foster the purchase of insurance are artificially maintaining an already depressed demand that could decline more dramatically in the future.

[40] Hofstede cultural proxies are related to the "Power distance index" (PDI), "Individualism versus Collectivism" (IDV), "Masculinity versus Feminity" (MAS), "Uncertainty avoidance index" (UAI), "Long-term Orientation versus Short-term Orientation" (Ltowvs), and "Indulgence Versus Restraint" (IVR).

[41] Related to the effect of research quality, we find that top-ranked journals seem to report smaller relative WTP.

The structural heterogeneity of our dataset allows us to define two different study profiles. The first one is related to survey-based studies with no information on the probability of loss, a subject pool representative of the population with no (small) participation fees. Such studies are expected to provide low RWTP. On the contrary, the second profile is related to studies with information about probabilities of losses, laboratory-based with participation fees, within-design, and young subjects with low income. For this second profile, we expect to observe a large RWTP. Future studies should strive to account for the observed WTP dependence on risk information and socioeconomic factors. Providing extremely low probability information seems to be more difficult to process and it is more prone to decision heuristics and cognitive bias (e.g. Kunreuther and Slovic, 1978; Hertwig et al., 2004; Kunreuther et al., 2001).

Future studies should also take heed of the sensitivity of stated WTP estimation to the qualitative characteristics of the subjects' sample in terms of randomness and representativity. A non-random sample may indirectly lead to biased WTP estimations as it may overrepresent some socioeconomic groups exacerbating the effect of age, income, or other moderators not considered in our study due to lack of data (e.g. loss experience, education, financial literacy, etc.).

---

[42] The overall average metric as an estimate of the true willingness to pay to insure LPHI risks is not different from the expected losses. This result may suggest no a priori global underestimation of tail losses nor a systematic rejection of insurance that might be considered by households, under a narrow framing context, as a poor financial investment (Gottlieb and Smetters, 2020). However, this result has to be viewed with caution and should be confirmed with further investigations including more populations. This result should however be interpreted with caution due to the small sample size or the possibility to be an artifact of econometric assumptions or RWTP metric definition. Future studies, however, will be needed to test whether average bias is associated with a null risk premium and to explore whether this finding is unique to the sample of studies under consideration.

**Table 4** Estimation of the multivariate meta-regression

This table reports the meta-regression results. The dependent variable of regression is the standardized willingness to pay (SWTP). Explanatory variables are defined in table 3. Three estimators are considered: (1) a cluster-robust ordinary least squares (OLS); (2) a cluster-robust random effects model (Unweighted RE) (3) a weighted random effects model by the inverse of the standard error (Weighted RE). The last column presents the regression results from BMA. For the BMA, the intercept posterior standard error is not available. As such, we recommend interpreting PIP value with caution.

| | (1)Unweighted OLS (Clustered) | | | (2) Unweighted RE (Clustered) | | | (3) Weighted RE | | | (4) Weighted BMA | | |
|---|---|---|---|---|---|---|---|---|---|---|---|---|
| | Coef. | SE | pval | Coef. | SE | pval | Coef. | SE | pval | Post Mean | Post SE | PIP |
| $\alpha_0$ (Precision) | -0.1957 | 0.6779 | 0.7742 | -0.1054 | 0.3942 | 0.7934 | -0.2304 | 0.2613 | 0.3828 | 0.3248 | 0.4102 | 0.4904 |
| $\alpha_1$ (Pub. bias) | 0.127 | 1.2851 | 0.9217 | -0.6998 | 1.3425 | 0.6109 | 0.0877 | 0.1381 | 0.5288 | -1.2171 | NA | 1 |
| Lab | **1.423*** | 0.2135 | **0.0001** | **1.4585*** | 0.1375 | **0.0001** | **1.3451*** | 0.1080 | **0.0000** | **1.3168** | **0.1533** | **1** |
| Online | 0.2322 | 0.2072 | 0.2688 | 0.1844 | 0.137 | 0.2014 | 0.1441 | 0.1486 | 0.3376 | 0.0870 | 0.1882 | 0.3522 |
| Within_design | **0.5433** | 0.1741 | **0.0033** | **0.5396** | 0.1481 | **0.003** | **0.4682*** | 0.1130 | **0.0002** | **0.4306** | **0.0829** | **1.0000** |
| Showup | **1.437*** | 0.2509 | **0.0001** | **1.5404*** | 0.2149 | **0.0001** | **1.5017*** | 0.1753 | **0.0000** | **1.1290** | **0.1600** | **1** |
| Incentive_compatible | -0.3602 | 0.2796 | 0.2047 | -0.3723* | 0.1975 | **0.082** | -0.4694** | 0.1628 | **0.0062** | -0.0344 | 0.1028 | 0.1967 |
| Elicit_method | 0.0797 | 0.1067 | 0.4595 | 0.0892 | 0.0841 | 0.3085 | 0.1403* | 0.0598 | **0.0239** | 0.0112 | 0.0357 | 0.2050 |
| Probability_level | **-27.966*** | 5.4187 | **0.0001** | **-33.225*** | 2.6467 | **0.0001** | **-27.930*** | 3.7179 | **0.0000** | **-25.7693** | **5.4770** | **0.9994** |
| Implicit_prob | -0.209 | 0.1645 | 0.211 | -0.2827* | 0.1376 | **0.0606** | -0.2384** | 0.0860 | **0.0083** | **-0.1255** | **0.1042** | **0.7103** |
| Risk_idiosync | 0.0395 | 0.2154 | 0.8553 | 0.1302 | 0.1393 | 0.3669 | 0.1198 | 0.0945 | 0.2118 | 0.0029 | 0.0683 | 0.1402 |
| Earthquake | **1.9783*** | 0.5103 | **0.0004** | **2.0919*** | 0.3638 | **0.0001** | **2.0016*** | 0.2621 | **0.0000** | **1.6628** | **0.3206** | **0.9999** |
| Sample size | **0.0005*** | 0.0002 | **0.0154** | **0.0003*** | 0.0001 | **0.0283** | **0.0002** | 0.0001 | **0.0024** | **0.0004** | **0.0001** | **0.9991** |
| China | **-1.3885*** | 0.1659 | **0.0001** | **-1.2488*** | 0.2285 | **0.0001** | **-1.1231*** | 0.1646 | **0.0000** | **-1.3690** | **0.1133** | **1** |
| Year | **-0.0761*** | 0.0132 | **0.0001** | **-0.0663*** | 0.0104 | **0.0001** | **-0.0658*** | 0.0063 | **0.0000** | **-0.0770** | **0.0078** | **1** |
| Germany | 0.3067 | 0.3127 | 0.3324 | 0.442 | 0.2595 | 0.1123 | 0.4737 | 0.2443 | 0.0592 | 0.0581 | 0.1455 | 0.2748 |
| Netherlands | **-0.6131*** | 0.2794 | **0.0338** | **-0.5004*** | 0.1764 | **0.014** | **-0.5173*** | 0.1350 | **0.0004** | **-0.3087** | **0.3141** | **0.6226** |
| Protest | 0.0735 | 0.1795 | 0.6842 | 0.1865 | 0.1796 | 0.318 | 0.1602 | 0.1306 | 0.2267 | 0.0348 | 0.0687 | 0.2944 |
| Random_sample | **0.5957** | 0.1808 | **0.002** | **0.7408*** | 0.1237 | **0.0001** | **0.6823*** | 0.1133 | **0.0000** | **0.5309** | **0.1940** | **0.9602** |
| Inverse_income | **0.3394*** | 0.0688 | **0.0001** | **0.2799*** | 0.0576 | **0.0003** | **0.2671*** | 0.0396 | **0.0000** | **0.3186** | **0.0644** | **0.9999** |
| Age | -0.0187 | 0.0131 | 0.1597 | **-0.0206*** | 0.0073 | **0.0146** | **-0.0194**** | 0.0057 | **0.0014** | **-0.0246** | **0.0071** | **0.9989** |
| Top_ranked | **-0.6102** | 0.1782 | **0.0014** | **-0.8085*** | 0.1245 | **0.0001** | **-0.6974*** | 0.0965 | **0.0000** | **-0.5202** | **0.1263** | **0.9976** |
| Low_citations | -0.0735 | 0.1405 | 0.6035 | -0.2988* | 0.1648 | **0.0929** | -0.3011* | 0.1244 | **0.0199** | 0.0020 | 0.0307 | 0.1237 |
| $R^2$ | | 94.02% | | | 91.53% | | | 91.53% | | | - | |
| $H^2$ | | - | | | 32.66 | | | 32.66 | | | - | |
| Q stat. (p.value) | | - | | | 449.67 (.0001) | | | 449.67 (.0001) | | | - | |
| Tau$^2$ (SE) | | - | | | 0.047 (0.0121) | | | 0.047 (0.0121) | | | - | |
| $I^2$ | | - | | | 96.94% | | | 96.94% | | | - | |
| N-obs | | 65 | | | 65 | | | 65 | | | 65 | |

**Notes:** Variables with PIP above 0.5 or significant are emphasized in bold. SD = standard deviation. SE = standard error. PIP= posterior inclusion probability. N.A. = not available. *, **, and *** denote statistical significance at 10%, 1%, and .1%, respectively. The last rows report regression $R^2$, $H^2$, Q, Tau$^2$ and $I^2$ statistics.

**Table 5** Estimation of the multivariate general-to-specific stepwise meta-regression

This table reports the meta-regression results based on a trimmed sample of 65 studies. The dependent variable of regression is the standardized willingness to pay (SWTP). Explanatory variables are defined in table 3. The results are obtained from a cluster-random-effects model: a weighted random effects model by the inverse of the standard error (Weighted RE). The last rows report regression $R^2$, $H^2$, Q, $Tau^2$ and $I^2$ statistics.

| | Model (1) | | | Model (2) | | | Model (3) | | |
|---|---|---|---|---|---|---|---|---|---|
| | Coef. | SE | pval | Coef. | SE | pval | Coef. | SE | pval |
| $\alpha_0$ (Precision) | -0.347* | 0.1375 | **0.0161** | 0.1425 | 0.5502 | 0.7978 | 0.4688 | 0.3297 | 0.1698 |
| $\alpha_1$ (Pub. bias) | - | - | - | -0.1527 | 1.0873 | 0.8895 | -1.3379 | 1.2567 | 0.2991 |
| Lab | - | - | - | **0.870*** | 0.2083 | **0.0003** | **1.1886*** | 0.1036 | **0.0001** |
| Online | - | - | - | - | - | - | - | - | - |
| Within_design | - | - | - | **0.3901*** | 0.1645 | **0.0257** | **0.4817*** | 0.1098 | **0.0003** |
| Showup | - | - | - | **1.254*** | 0.3285 | **0.0008** | **1.3387*** | 0.1958 | **0.0001** |
| Incentive_compatible | - | - | - | - | - | - | - | - | - |
| Elicit_method | - | - | - | - | - | - | - | - | - |
| Probability_level | - | - | - | **-23.09*** | 7.5271 | **0.0051** | **-31.581*** | 3.7465 | **0.0001** |
| Implicit_prob | - | - | - | - | - | - | -0.3182* | 0.1404 | **0.0341** |
| Risk_idiosync | - | - | - | - | - | - | - | - | - |
| Earthquake | - | - | - | - | - | - | **1.8898*** | 0.2432 | **0.0001** |
| Sample size | - | - | - | - | - | - | - | - | - |
| China | - | - | - | **-0.917*** | 0.2081 | **0.0002** | **-1.3709*** | 0.1692 | **0.0001** |
| Year | - | - | - | **-0.044*** | 0.0135 | **0.0028** | **-0.0777*** | 0.0118 | **0.0001** |
| Germany | - | - | - | - | - | - | - | - | - |
| Netherlands | - | - | - | -0.488* | 0.2075 | **0.0268** | **-0.4425*** | 0.1402 | **0.0048** |
| Protest | - | - | - | - | - | - | - | - | - |
| Random_sample | - | - | - | 0.3013 | 0.3594 | 0.4097 | **0.716*** | 0.1474 | **0.0001** |
| Inverse_income | - | - | - | - | - | - | **0.2116*** | 0.0733 | **0.0088** |
| Age | - | - | - | **-0.0173*** | 0.006 | **0.0078** | **-0.0235*** | 0.0035 | **0.0001** |
| Top_ranked | - | - | - | - | - | - | **-0.6104*** | 0.1441 | **0.0004** |
| Low_citations | - | - | - | - | - | - | - | - | - |
| $R^2$ | | 0.00% | | | 71.20% | | | 88.49% | |
| $H^2$ | | 410.42 | | | 108.63 | | | 43.27 | |
| Q stat (p.value) | | 16616.53 (.0001) | | | 1847.98 (.0001) | | | 858 (.0001) | |
| $Tau^2$ (SE) | | 0.560 (0.1014) | | | 0.1613 (0.0329) | | | 0.0645 (0.0145) | |
| $I^2$ | | 99.76% | | | 99.08% | | | 97.69% | |
| N-obs | | 65 | | | 65 | | | 65 | |

**Notes:** *, **, and *** denote statistical significance at 10, 1%, and .1%, respectively.

**5.4. Robustness checks**

We estimate additional regression models to check the robustness of the results obtained in Table 4. First, we perform BMA analysis with a weight equal to the inverse of the number of data points per study and then without weighting. The results given in Table 6 corroborate previous findings, where the magnitude and the sign of the moderators exhibit little variation.

To further investigate heterogeneity, we apply a three-level structure to the meta-regression model which allows for examining differences in outcomes within studies (i.e., within-study heterogeneity) as well as differences between studies (i.e., between-study heterogeneity).[43] Unlike other models, we do not need to know correlations between outcomes reported within primary studies since the second level accounts for sampling covariation (Van den Noortgate et al., 2013). Note that the random effects hierarchical method that we use for estimation allows coefficients to vary randomly across studies (Ugur et al. 2016; Neves and Sequeira, 2018). We test different candidate variables as third level such as article Id, country, survey-based, risk type, and coverage type. The results do not change from previous findings, confirming the absence of correlation between SWTP within studies and therefore the appropriateness of the two-level model used to analyze heterogeneity.[44] None of the third-level candidate variables can explain more of the variability between studies.

As a fourth robustness check, we conduct outlier analyses by first examining extreme SWTP with confidence intervals that did not overlap with the confidence interval of the pooled effect. We perform influence analyses via a "leave-one-out" method, in which effect size is recalculated when a single study is left out of the analysis (Viechtbauer and Cheung, 2010). We identified three observations as potential outliers or influential outcomes with the leave-one-out analysis and the Baujat plot displayed in Figure 7. After progressively excluding these three data points, we further reduce heterogeneity while confirming the obtained estimation results.[45] Finally, as a last robustness check we estimate a meta-regression model using conditional SWTP (excluding zeros WTP) as effect size. The estimation results are quite similar to those reported in Table 4.[46]

*Figure* 7: Baujat plot of SWTP

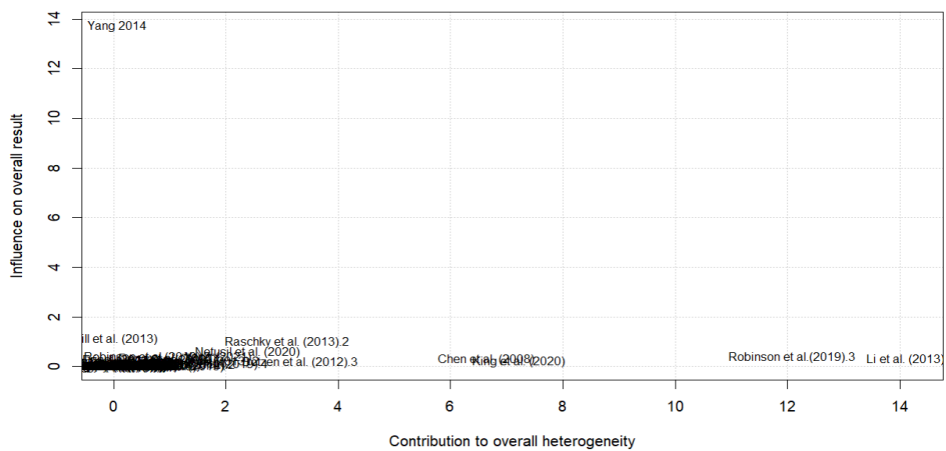

The *X*-axis represents the contribution of each WTP estimate to the overall Cochran *Q*-test for heterogeneity. The *Y*-axis represents the influence of WTP estimates on the overall average level.

**Table 6** Additional BMA meta-regression

---

[43] The three-level meta-analytic model assumes different variance components distributed over the levels of the model: sampling variance of the SWTP extracted at level 1; variance between effect sizes extracted from the same study at level 2; and variance between studies at level 3.
[44] Estimation results are reported in Appendix I.
[45] Estimation results are reported in Appendix J.
[46] Estimation results are reported in Appendix K.

This table reports additional meta-regression results from BMA without weighting and with a weight equal to the inverse of the number of data points per study. The dependent variable of regression is the standardized willingness to pay (SWTP). Explanatory variables are defined in Table 3. For the precision (publication bias) variable, the posterior standard error in the unweighted(weighted) model is not available. As such, we recommend interpreting PIP values with caution.

|  | Unweighted BMA | | | Weighted BMA (number of data points per study) | | |
| --- | --- | --- | --- | --- | --- | --- |
|  | Post Mean | Post SE | PIP | Post Mean | Post SE | PIP |
| $\alpha_0$ (Precision) | 0.1469 | NA | 1 | -0.0004 | 0.1541 | 0.14031 |
| $\alpha_1$ (Pub. bias) | - | - | - | -0.0195 | NA | 1 |
| Std. error | -0.0160 | 0.1307 | 0.1528 | -0.0167 | 0.0806 | 0.1492 |
| **Lab** | **1.2467** | **0.2291** | **1** | **1.2161** | **0.1587** | **1** |
| Online | 0.0752 | 0.1706 | 0.2788 | 0.0709 | 0.1827 | 0.2490 |
| **Within_design** | **0.4964** | **0.1464** | **0.9870** | **0.5225** | **0.1590** | **0.9873** |
| **Showup** | **1.3927** | **0.1893** | **1** | **1.2912** | **0.1477** | **1** |
| Incentive_compatible | -0.0370 | 0.1408 | 0.1834 | -0.0838 | 0.1527 | 0.3263 |
| Elicit_method | 0.0078 | 0.0425 | 0.1615 | 0.0008 | 0.0271 | 0.1381 |
| **Probability_level** | **-33.928** | **4.6663** | **1** | **-31.8202** | **4.8846** | **1** |
| **Implicit_prob** | **-0.3060** | **0.1923** | **0.8160** | **-0.4661** | **0.0982** | **0.9996** |
| Idiosync_risk | 0.0315 | 0.1245 | 0.1654 | 0.0103 | 0.0662 | 0.1285 |
| **Earthquake** | **1.9184** | **0.4919** | **0.9925** | **2.0572** | **0.2786** | **1** |
| Sample_size | 0.0001 | 0.0001 | 0.4193 | 0.0000 | 0.0001 | 0.2633 |
| **China** | **-1.2403** | **0.1636** | **1** | **-1.2151** | **0.0995** | **1** |
| **Year** | **-0.0695** | **0.0147** | **0.9998** | **-0.0736** | **0.0105** | **1** |
| Germany | 0.1960 | 0.2563 | 0.4812 | **0.2546** | **0.2367** | **0.6421** |
| **Netherlands** | **-0.5309** | **0.2450** | **0.9306** | **-0.4813** | **0.2116** | **0.9498** |
| Protest | 0.1178 | 0.1552 | 0.4699 | 0.0243 | 0.0637 | 0.2179 |
| **Random_sample** | **0.7529** | **0.2169** | **0.9863** | **0.7594** | **0.1458** | **0.9998** |
| **Inverse_income** | **0.1916** | **0.1087** | **0.8539** | **0.2534** | **0.0518** | **0.9998** |
| **Age** | **-0.0226** | **0.0101** | **0.9283** | **-0.0164** | **0.0045** | **0.9859** |
| **Top_ranked** | **-0.6349** | **0.1480** | **0.9993** | **-0.6617** | **0.1208** | **0.9999** |
| **Low_citations** | **-0.1513** | **0.1619** | **0.5712** | **-0.1209** | **0.1233** | **0.5907** |

Notes: Variables with PIP above 0.5 are shown in bold. SD = standard deviation. SE = standard error. PIP= posterior inclusion probability. N.A. = not available. Number of observations: 65.

## 6. Conclusion

We present a meta-analysis of contingent valuation studies for low-probability risk insurance. Consistent with the few observed market-based data, the meta-analytic average willingness to pay found is lower than expected losses. Survey-based studies exhibited a particularly low weighted mean of 64%. Normalized WTP levels vary considerably across studies allowing meta-regression capturing different sources of heterogeneity. The main finding is that the variability of stated WTP is structurally dependent on risk characteristics, fundamental and methodological factors. Some of these factors have a strong theoretical basis to explain low insurance demand at the individual level.

Our results provide evidence of exogenous determinants that may affect the relative WTP variability at the study level. Moderators such as information about probabilities and very small probability levels appear to positively influence relative WTP, whereas some respondents' sociodemographic characteristics such as income and age show a negative effect. Laboratory-based estimates, appear also to report significantly higher values for relative WTP than alternative data collection methods. These factors are likely to accentuate cognitive biases and judgment errors, particularly in the presence of several WTP elicitation tasks. Similarly, cultural factors related to power distance and uncertainty avoidance bring further explanations for discrepancies in insurance WTP across international samples. Our results also document a consistent downward trend of average WTPs over time, a finding that may be indicative of an increasing inattention to low-probability risks driven in part by the increasing acceptance of climate change perspectives and physical risk damages, especially in the absence of comprehensive information on the probability of losses.

To ensure the long-term viability of non-mandatory coverage, outreach efforts are required through insurance education and promotion actions to guarantee a sufficient pool of policyholders, taking mandatory insurance systems with very low premiums and no adverse selection as a boundary model. Furthermore, policymakers should continue to support individual vulnerability reduction measures, which seem to be effective in reducing the severity of losses and less subject to individuals' biases and heuristics. However, achieving these two essential objectives may prove more difficult for certain social groups with low financial capacity, for whom insurance may seem the primary safety net. One way to ease the budget constraint is to combine insurance mechanism with other flexible financial instruments such as credit access. For example, access to emergency loans may reduce the initial cash payment of premiums and lessen the liquidity constraint. Such a solution, consistent with the principle of discontinuity of preferences, would reflect the preference for more flexible risk management vehicles.

For non-price factors affecting demand, we still need more empirical evidence on the interactions between risk preferences and behavior biases, on the one hand, and insurance characteristics on the other. We can identify two avenues for future investigation. The first direction is to explore the interactions between risk preferences, insurance characteristics and decision complexity. This raises the question of whether insurance uptake should be studied alone or whether it is better seen as one facet of a complex risk management problem, as opposed to much of the literature. A second research avenue is on how cognitive and behavioral biases related to low-probability risks and insurance purchase may persist over time and how they can be exacerbated by some public policies. A better understanding of these intermingling dynamics will provide valuable policy guidance, through more efficient subsidy targeting and/or a progressive shift toward risk-based pricing systems. Overall, our results provide additional credit for contingent valuation methods and alleviate some concerns about their external validity.[47]

Our study is not without limitations, although it passes most of the robustness checks. We acknowledge the dataset size restriction and its potential impact on the stability of the meta-regression coefficient. This constraint also limits our ability to test additional non-linear and interaction effects between moderators to further explain the heterogeneity. The second limitation is that we were unable to explore potentially significant latent drivers of WTP due to a lack of data. Some important factors, such as financial literacy, past loss experience, or perceived insurance providers' quality are not included in the meta-regression analysis. Third, because protesters are not systematically identified and corrected across studies, false-zero WTP responses may occur, leading to a systematic downward bias in the results. In the absence of market data, stated preference methods would offer an effective method to measure insurance demand. While our results offer no direct support for hypothetical bias, the effect of methodology-related aspects identified in this paper should be acknowledged in the design of future stated preference measurements. In light of what is at stake, these studies should attempt to resolve some of these issues.

---

[47] A "Dynamic" Meta-Analysis project for WTP for insurance against LPHI risks is under development (see Appendix L for further details).

**Meta-Analysis References**

# Supplementary material

**Appendix A: Derivation of SWTP and var (SWTP) sampling characteristics:**

The following steps are inspired from Lajeunesse (2015). The $SWTP$ metric and its variance $\text{var}(SWTP)$ can be formulated as first-order approximations of the log ratio of a random variable and a scalar. Following Stuart and Ord (1994), the expectation of the true level $\lambda = \ln\left(\dfrac{\mu}{EL}\right)$ estimator is based on the first-order Taylor expansion around the WTP population mean $\mu$ and expected loss (EL):

$$E(SWTP) \approx \lambda + \mathbf{J}^T(\mathbf{x} - \boldsymbol{\mu}) + \varepsilon_{SWTP} \qquad (A.1)$$

where the superscript $T$ indicates the transposition of a matrix, $\varepsilon_{SWTP}$ is the ignored higher-order Taylor expansions, $\boldsymbol{\mu}$ a column vector of the population mean $\mu$ and $EL$ i.e. $\boldsymbol{\mu}^T = [\mu, EL]$ and $\mathbf{x}$ a vector of the sample means i.e. $\mathbf{x}^T = [\overline{WTP}, \overline{EL}]$. The Jacobian vector ($\mathbf{J}$) containing all the first-order partial derivatives ($\partial$) of each variable in $\lambda$ is:

$$\mathbf{J}^T = \left[\dfrac{\partial \lambda}{\partial \mu} \quad \dfrac{\partial \lambda}{\partial (EL)}\right] = \left[\dfrac{1}{\mu} \quad -\dfrac{1}{EL}\right] \qquad (A.2)$$

Note that the expectation of $\left(\overline{WTP} - \mu\right)$ tends to zero at large sample sizes, (Stuart and Ord 1994). We get the original formulation of the SWTP:

$$E(SWTP) \approx \ln\left(\dfrac{\mu}{EL}\right) + \dfrac{\overline{WTP} - \mu}{\mu} - \dfrac{\overline{EL} - EL}{EL} \approx \ln\left(\dfrac{\mu}{EL}\right) \approx \lambda \qquad (A.3)$$

The approximated variance of SWTP is:

$$\text{var}(SWTP) \approx \mathbf{J}^T \mathbf{V} \mathbf{J} + \varepsilon_{SWTP} \qquad (A.4)$$

Where $\mathbf{V}$ is the variance–covariance matrix of $\mu$ and $EL$ containing their large-sample variances and zero covariances as follows:

$$V = \begin{bmatrix} \dfrac{\sigma^2}{N} & 0 \\ 0 & 0 \end{bmatrix}$$

Where N is the sample size. By solving Eq. 4, we get the variance:

$$\text{var}(SWTP) \approx \dfrac{\sigma^2}{N\mu^2} + 0 \qquad (A.5)$$

When replacing the population parameters $\mu$ and $\sigma^2$ with their respective sample statistics, $\overline{WTP}$ and SD², we get equations 3 and 4 given in the main text.

**Appendix B:** Keyword combinations and Google Scholar results

| | Keywords combination | Number of documents |
|---|---|---|

|  |  | Google Scholar |
|---|---|---|
| Set1 | Insurance "Willingness to pay" "low probability» -intitle:Health | 4124 |
| Set2 | Insurance "willingness to pay" "low probabilities" -intitle:Health | 832 |
| Set3 | "Flood risk" insurance "willingness to pay"  -intitle:Health | 4342 |
| Set4 | "Climate risk" Insurance "Willingness to pay"  -intitle:Health | 2600 |
| Set5 | "Natural disasters" Insurance "Willingness to pay" -intitle:Health | 7064 |
| Set6 | "Contingent valuation" Insurance "Willingness to pay" -intitle:Health | 11011 |
| Set7 | Kunreuther Insurance "Willingness to pay"  -intitle:Health | 1953 |

**Notes**: Documents' search on Google Scholar database over the period 2005-2021. After merging results from different search sets and removing duplicates, we have a total of 15,664 documents.

**Appendix C:** List of primary studies in the dataset and expected loss sources

This table lists the information of primary studies included in the dataset (n=65). The four first columns report general information (e.g., article id, title, first author, and number of studies). The last column shows the method for estimating the expected loss (EL). This variable can be either (1) estimated from historical average losses, (2) calculated from known loss distributions (as prob x loss), or finally (3) measured from the provided information on the actuarially fair premium.

| Article ID | Title | First Author | Number of studies | Expected loss |
|---|---|---|---|---|
| 1 | An experimental investigation of insurance decisions in low probability and high loss risk situations | Ozedmir et al. (2013) | 4 | Calculated as prob x Loss |
| 2 | An incentive-compatible experiment on probabilistic insurance and implications for an insurer's solvency level | Zimmer et al. (2016) | 1 | Calculated as prob x Loss |
| 3 | Feeling the numbers: On the Interplay between risk, affect, and numeracy | Petrova et al. (2013) | 2 | Calculated as prob x Loss |
| 4 | Comparing the effects of behaviorally informed interventions on flood insurance demand: an experimental analysis of 'boosts' and 'nudges' | Bradt (2019) | 1 | Calculated as prob x Loss |
| 5 | Flood risk perceptions and the willingness to pay for flood insurance in the Veneto region of Italy | Roder et al. (2019) | 1 | Premium value available |
| 6 | Risk attitudes to low-probability climate change risks: WTP for flood insurance | Botzen et al. (2012) | 3 | Calculated as prob x Loss |
| 7 | Behavioral motivations for self-insurance under different disaster risk insurance schemes | Mol et al. (2020) | 1 | Calculated as prob x Loss |
| 8 | What drives the willingness to pay for insurance contracts with nonperformance risk? Experimental evidence | Hillebrandt (2020) | 3 | Calculated as prob x Loss |
| 9 | Catastrophic risk: social influences on insurance decisions | Krawczyk et al. (2017) | 2 | calculated (Average prob x loss) |
| 10 | Seismic risk-coping behavior in rural ethnic minority communities in Dali, China | Zhang (2020) | 1 | Premium value available |
| 11 | Determinants of Probability Neglect and Risk Attitudes for Disaster Risk: An Online Experimental Study of Flood Insurance Demand among Homeowners | Robinson et al. (2019) | 8 | Calculated as prob x Loss |
| 12 | Household Preference and Financial Commitment to Flood Insurance in South-East Queensland | Lo (2013) | 1 | Premium value available |
| 13 | The Willingness to Pay for Flood Insurance | Netusil et al. (2020) | 1 | Premium value available |
| 14 | Rural homeowners' willingness to buy flood insurance | Ren et al. (2016) | 3 | Premium value available |
| 15 | Uncertainty of Governmental Relief and the Crowding out of Flood Insurance | Raschky et al. (2013) | 2 | (1) Adjusted average loss (2) Adapted from Seifert et al.(2013) |
| 16 | Influence of flood risk characteristics on flood insurance demand: a comparison between Germany and the Netherlands | Seifert et al. (2013) | 2 | (1) Calculated as prob x Loss (2) Average loss |
| 17 | Farmers' Willingness to Pay for Cow Insurance in Shaanxi Province, China | Xiu et al. (2012) | 1 | Premium value available |
| 18 | Is default risk acceptable when purchasing insurance? Experimental evidence for different probability representations, reasons for default, and framings | Zimmer et al. (2009) | 6 | Premium value available |
| 19 | Factors influencing Shaanxi and Gansu farmers' willingness to purchase weather insurance | Kong et al. (2011) | 1 | Calculated as prob x Loss |
| 20 | Understanding farmers' valuation of agricultural insurance: Evidence from Vietnam | King et al. (2020) | 1 | Calculated as prob x Loss |
| 21 | Certain and Uncertain Utility and Insurance Demand: Results From a Framed Field Experiment in Burkina Faso | Seafilippi et al. (2020) | 1 | Calculated as prob x Loss |
| 22 | Adoption of weather-index insurance: learning from willingness to pay among a panel of households in rural Ethiopia | Hill et al. (2013) | 1 | Premium value available |
| 23 | Willingness to Pay For Index Based Crop Insurance In Ghana | Ellis (2017) | 2 | Premium value available |
| 24 | Productivity, credit, risk, and the demand for weather index insurance in smallholder agriculture in Ethiopia | McIntosh et al. (2013) | 1 | Premium value available |
| 25 | Insurance demand and social comparison: An experimental analysis | Friedl et al. (2014) | 2 | Calculated as prob x Loss |
| 26 | Evaluation of the crop insurance management for soybean risk of natural disasters in Jilin Province, China | Yang et al. (2015) | 1 | Premium value available |
| 27 | Crop insurance knowledge, trust in government and demand for crop insurance-an empirical study of peasant households' willingness-to-pay in Huaian, Jiangsu Province | Sun (2008) | 6 | Premium value available |
| 28 | Study on Farmers' Willingness to Pay for Policy Forest Insurance Based on Cox Model | Li et al. (2013) | 3 | Premium value available |
| 29 | An Empirical Analysis of Farmers' Willingness to Pay for Agricultural Insurance Take the Manas River Basin in Xinjiang as an example | Ning et al. (2006) | 2 | Premium value available |
| 30 | Analysis of Farmers' Willingness to Pay for Agricultural Insurance and Its Influencing Factors Take tobacco insurance in Xingshan County, Hubei Province as an example | Chen et al. (2008) | 1 | Premium value available |
| 31 | Exploring farmers' willingness to pay for crop insurance products: A case of weather-based crop insurance in Punjab, India | Aditya et al. (2020) | 1 | Premium value available |
| 32 | Farmers' interest and willingness-to-pay for index-based crop insurance in the lowlands of Nepal | Budhathoki et al. (2019) | 2 | Premium value available |
| 33 | Willingness-to-pay for yak snow disaster weather index insurance: A case study of Yushu Tibetan Autonomous Prefecture, Qinghai Province | Yang et al. (2021) | 1 | Premium value available |
| 34 | Climate perceptions, farmers' willingness-to-insure farms and resilience to climate change in Northern region, Ghana | Adzawla et al. (2019) | 1 | Premium value available |
| 35 | Estimating farmers' willingness to pay for weather index-based crop insurance uptake in West Africa: Insight from a pilot initiative in Southwestern Burkina Faso | Fonta et al. (2018) | 1 | Premium value available |
| 36 | The willingness of farmers to pay insurance premiums for sustainable rice farming in Bali | Budiasa et al. (2020) | 1 | Premium value available |
| 37 | Willingness to pay for potential standing timber insurance | Deng et al. (2015) | 1 | Premium value available |
| 38 | Evaluating the demand for aquaculture insurance: An investigation of fish farmers' WTP in central coastal areas in China | Zheng (2018) | 1 | Average loss available |

## Appendix D: Distribution of studies

These figures show the relative frequencies of in included studies (n=65) according to different subgroups (e.g., continent, year, country, coverage type, elicitation method, insurance type, risk type, and number of studies).

**Figure D.1:** Distribution of studies by continent

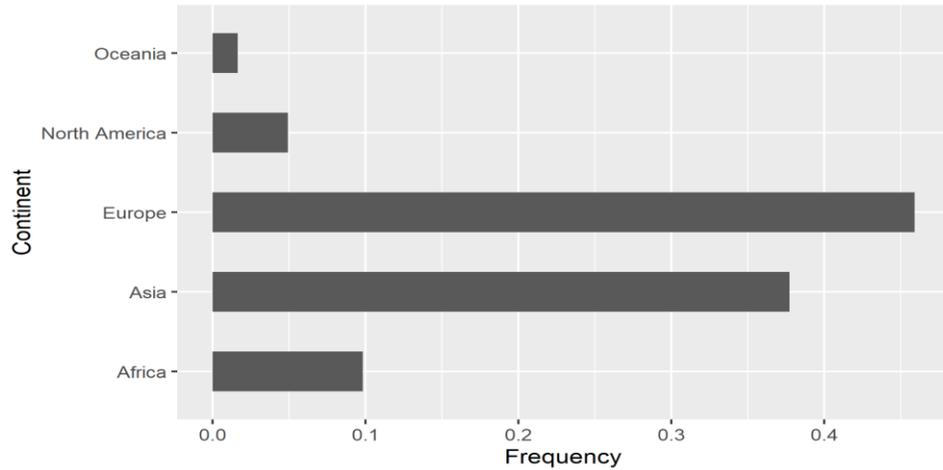

**Figure D.2:** Distribution of studies per year

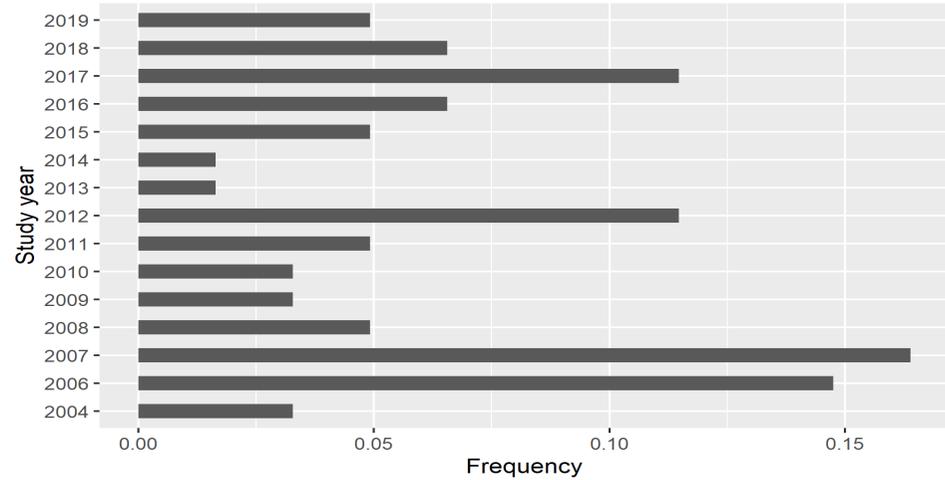

**Figure D.3:** Distribution of studies by country

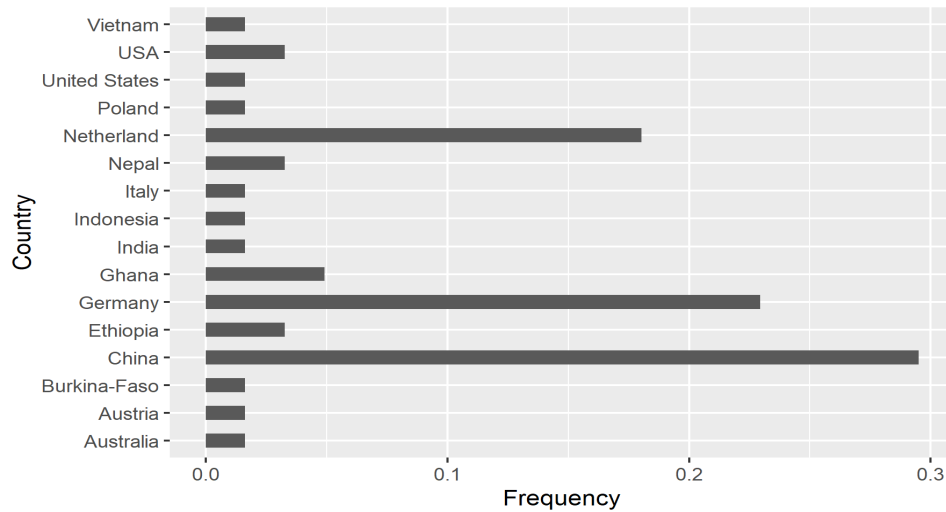

**Figure D.4:** Distribution of studies by coverage type

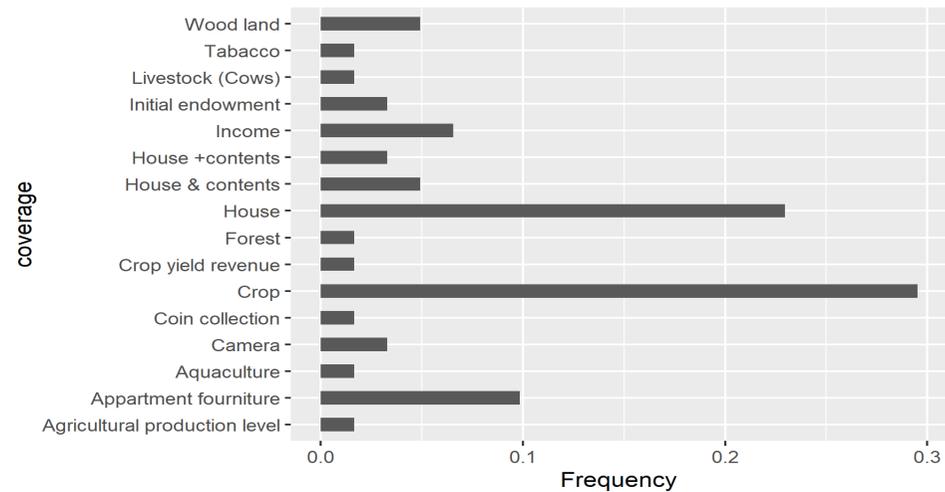

**Figure D.5**: Distribution of studies by elicitation method

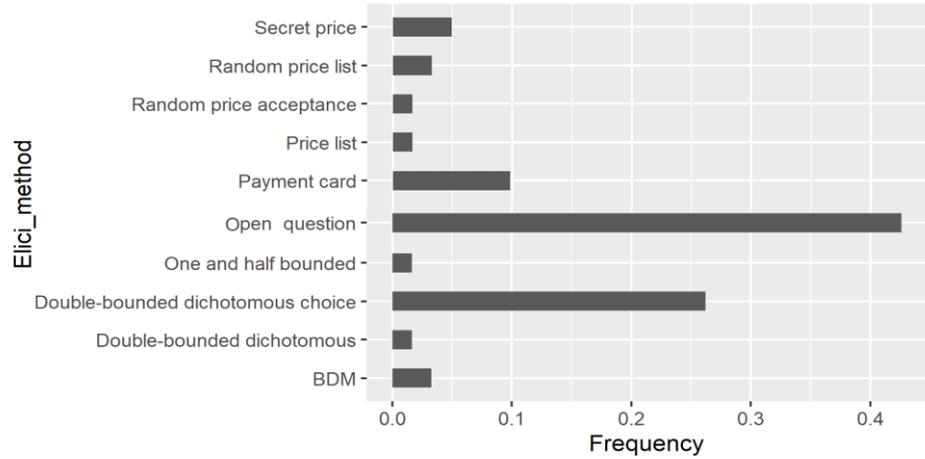

**Figure D.6**: Distribution of studies by insurance type

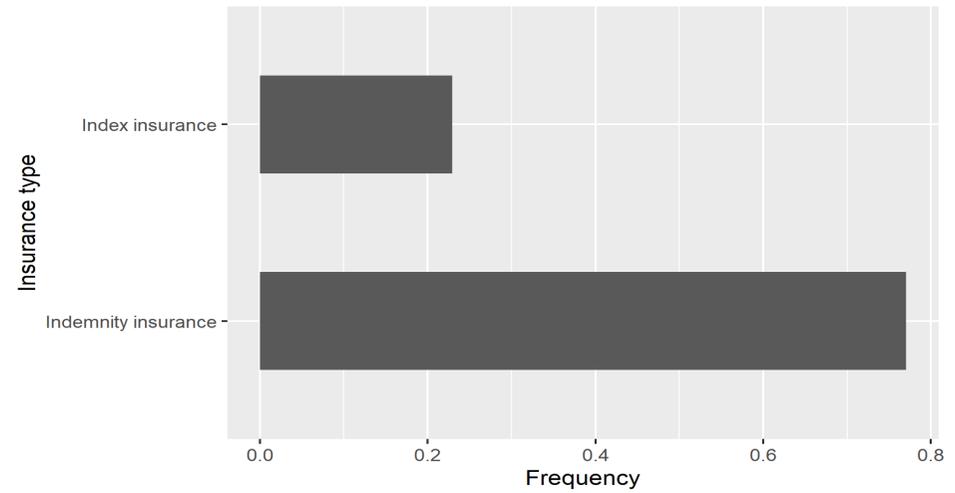

**Figure D.7**: Distribution of studies by risk type

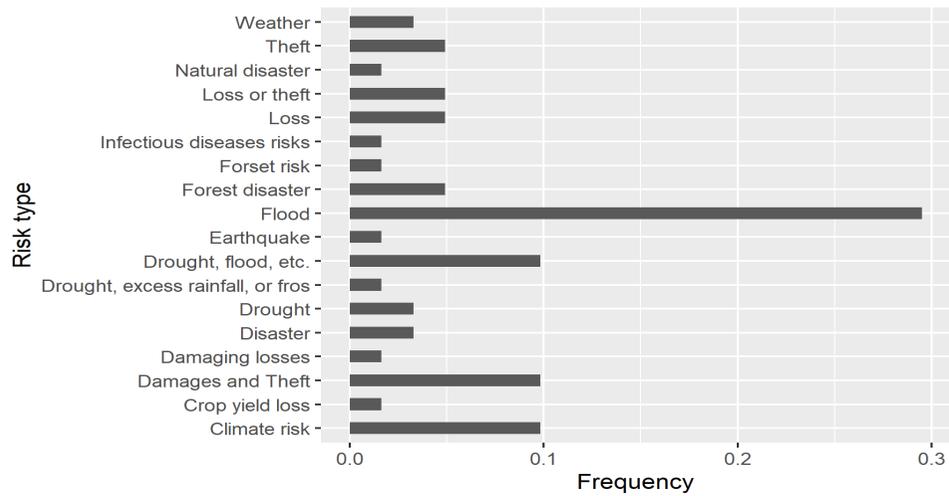

**Figure D.8**: Distribution by number of studies

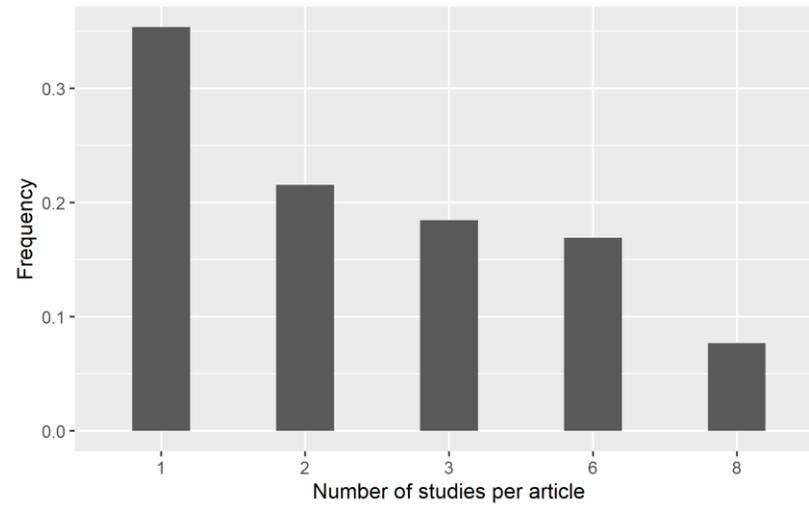

**Appendix E:** Overall average SWTP and Heterogeneity test

This table shows the overall average SWTP estimation and different heterogeneity indicators (Q-stat, H², I², Tau).

| Estimate | Std. error | t | P-value | ci.lb | ci.ub |
|---|---|---|---|---|---|
| -0.3555 | 0.1864 | -1.9075 | 0.0609 | -0.7279 | 0.0168 |

| Q-stat(df = 64) | P-value | H² | I² | Tau² | Tau |
|---|---|---|---|---|---|
| 16616.53 | .0001 | 410.42 | 99.76% | 0.56 (0.1) | 0.7484 |

**Appendix F:** Multi-collinearity test

This table shows the variance inflation factor (VIF) of moderatos included in the meta-regression

| Variables | Tolerance | VIF |
|---|---|---|
| Germany | 0.136 | 7.355 |
| Showup | 0.141 | 7.079 |
| Online | 0.142 | 7.021 |
| Risk_idiosync | 0.143 | 7.003 |
| Incentive_compatible1 | 0.150 | 6.645 |
| Netherlands | 0.151 | 6.612 |
| Elicit_method | 0.196 | 5.096 |
| Lab | 0.215 | 4.659 |
| China | 0.215 | 4.653 |
| Top_ranked | 0.218 | 4.591 |
| Age | 0.236 | 4.233 |
| Implicit_prob | 0.259 | 3.862 |
| Probability_level | 0.261 | 3.838 |
| Protest | 0.267 | 3.744 |
| Year | 0.296 | 3.381 |
| Within_design | 0.318 | 3.149 |
| Low_citations | 0.347 | 2.885 |
| Earthquake | 0.385 | 2.600 |
| Random_sample | 0.448 | 2.233 |
| Sample_size | 0.536 | 1.867 |
| std_errorlnwtp | 0.597 | 1.674 |
| Inverse_income | 0.689 | 1.451 |

**Appendix G**

**Figure G.1**: Forest Plot of relative WTP

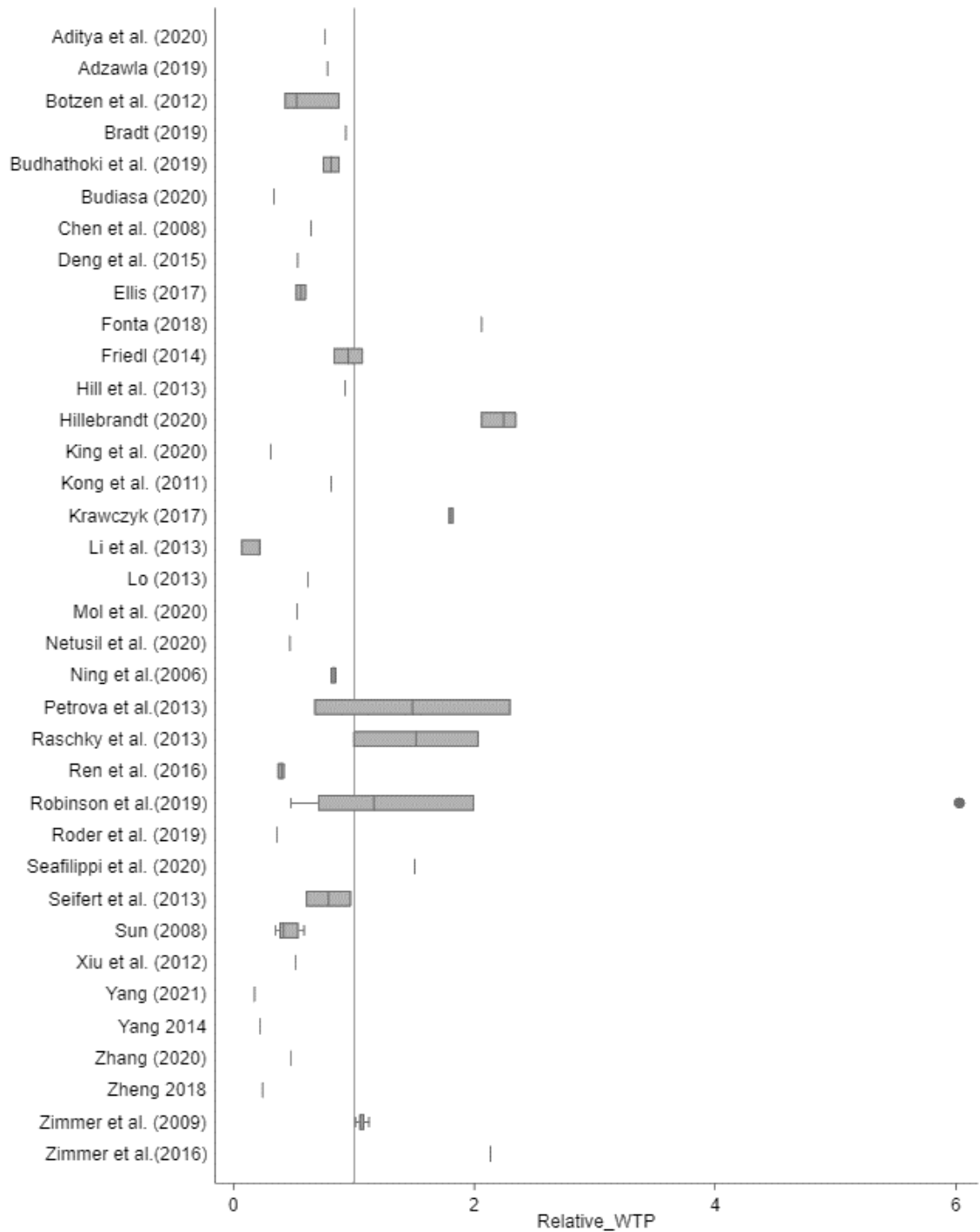

This figure shows the forest plot and summary RWPT for all studies (n = 65). The random-effects model was chosen because of the high heterogeneity induced by the different studies.

**Appendix H:** Visual Representation of BMA-meta regression including cultural factors (weighted by the inverse of SE)

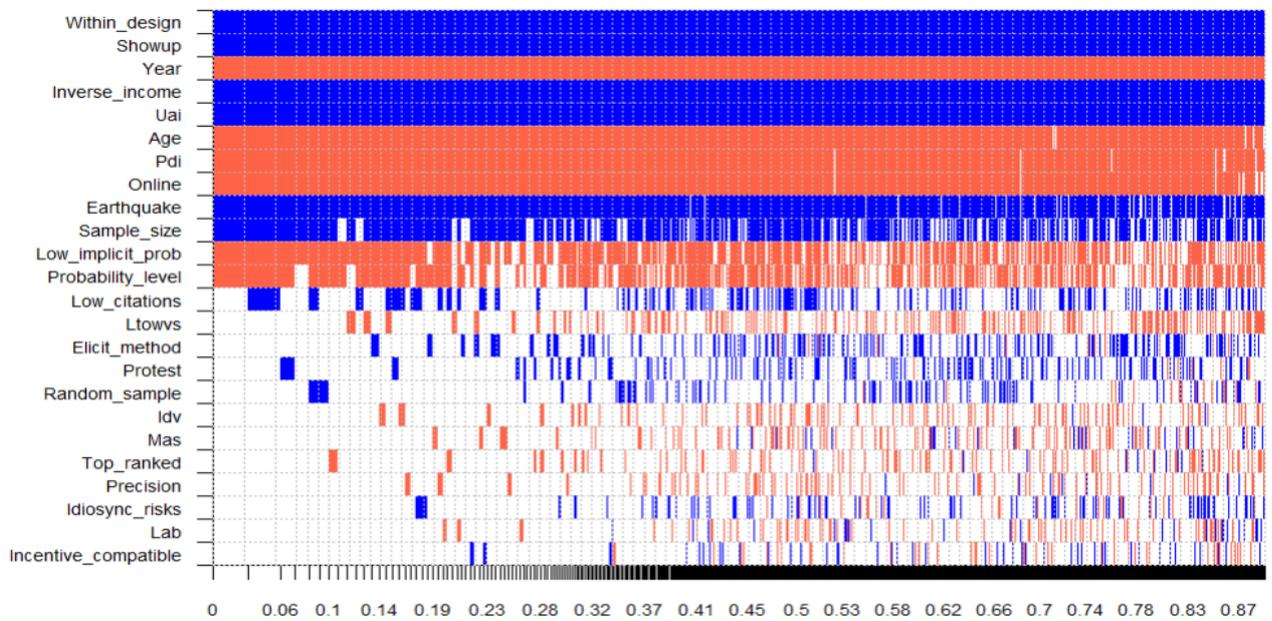

This figure shows the graphical result of meta regression using the Bayesian model averaging. All variables are described in Table 3. Hofstede cultural proxies are related to: power distance (PDI), collectivism (COL), masculinity (MAS), uncertainty avoidance (UAI), and long-term orientation (LTOWVS). On the vertical axis, the explanatory variables are ranked according to their posterior inclusion probabilities from the highest at the top to the lowest at the bottom. The horizontal axis shows the values of cumulative posterior model probability. Blue color (darker in grayscale) = the estimated parameter of a corresponding explanatory variable is positive. Red color (lighter in grayscale) = the estimated parameter of a corresponding explanatory variable is negative. No color = the corresponding explanatory variable is not included in the model.

**Appendix I:** BMA meta-regression including cultural factors (weighted by the inverse of SE)

This table reports the BMA meta regression results (weighted by the inverse of SE) including cultural factors. The dependent variable of regression is the standardized willingness to pay (SWTP). Hofstede cultural proxies are related to: power distance (PDI), collectivism (COL), masculinity (MAS), uncertainty avoidance (UAI), and long-term orientation (LTOWVS). Explanatory variables are defined in table 3. For the publication bias variable, the posterior standard error is not available. As such, we recommend interpreting PIP result with caution.

|  | Weighted BMA | | |
| --- | --- | --- | --- |
|  | Post Mean | Post SE | PIP |
| $\alpha_0$ (Precision) | -0.0472 | 0.2282 | 0.1395 |
| $\alpha_1$ (Pub. bias) | -1.5694 | NA | 1 |
| **Within_design** | **0.5370** | **0.0909** | **1** |
| **Showup** | **0.7880** | **0.1607** | **1** |
| **Year** | **-0.0512** | **0.0095** | **1** |
| **Inverse_income** | **0.3900** | **0.0772** | **1.0000** |
| **Uai** | **0.0301** | **0.0033** | **0.9999** |
| **Age** | **-0.0240** | **0.0059** | **0.9943** |
| **Pdi** | **-0.0154** | **0.0042** | **0.9875** |
| **Online** | **-0.6447** | **0.1708** | **0.9864** |
| **Earthquake** | **0.9651** | **0.3584** | **0.9636** |
| **Sample_size** | **0.0002** | **0.0002** | **0.7367** |
| **Implicit_prob** | **-0.1380** | **0.1064** | **0.7292** |
| **Probability_level** | **-7.9533** | **7.3503** | **0.6405** |
| Low_citations | 0.0423 | 0.0802 | 0.3038 |
| Ltowvs | -0.0007 | 0.0017 | 0.2316 |
| Elicit_method | 0.0165 | 0.0469 | 0.2110 |
| Protest | 0.0182 | 0.0567 | 0.1828 |
| Random_sample | 0.0225 | 0.1018 | 0.1530 |
| Idv | -0.0004 | 0.0018 | 0.1499 |
| Mas | -0.0003 | 0.0017 | 0.1461 |
| Top_ranked | -0.0154 | 0.0646 | 0.1437 |
| Idiosync_risks | 0.0131 | 0.0734 | 0.1346 |
| Lab | -0.0059 | 0.0972 | 0.1271 |
| Incentive_compatible | -0.0020 | 0.0614 | 0.1056 |

Notes: Variables with PIP above 0.5 are shown in bold. SD = standard deviation. SE = standard error. PIP= posterior inclusion probability. N.A. = not available. Number of observations: 65.

**Table I.**1: Results of hierarchical model estimation (article Id as third level)

This table reports the meta-regression results using the random effects hierarchical method considering article Id as third level. The dependent variable of regression is the standardized willingness to pay (SWTP). Explanatory variables are defined in Table 3.

|  | W-RE | | | RE | | |
|---|---|---|---|---|---|---|
|  | Coeff. | SE | pval | Coeff. | se | pval |
| $\alpha_0$ (Precision) | -0.1223 | 0.6597 | 0.8539 | -0.1054 | 0.373 | 0.7789 |
| **Lab** | 1.5157*** | 0.2608 | **0.0001** | 1.4585*** | 0.1867 | **0.0001** |
| **Online** | 0.2243 | 0.2548 | 0.3839 | 0.1844 | 0.1808 | 0.3135 |
| **Within_design** | 0.4381** | 0.1812 | **0.0203** | 0.5396*** | 0.1099 | **0.0001** |
| **Showup** | 1.5055*** | 0.2696 | **0.0001** | 1.5404*** | 0.1885 | **0.0001** |
| **Incentive_compatible1** | -0.3939 | 0.2861 | 0.1761 | -0.3723. | 0.1983 | 0.0674 |
| **Elicit_method** | 0.1123 | 0.1253 | 0.3753 | 0.0892 | 0.0769 | 0.253 |
| **Probability_level** | -26.964*** | 5.5102 | **0.0001** | -33.2259*** | 3.9707 | **0.0001** |
| **Implicit_prob** | -0.0217 | 0.1943 | 0.9114 | -0.2827* | 0.1157 | **0.0188** |
| **risk_idiosync** | 0.0631 | 0.1768 | 0.7233 | 0.1302 | 0.1828 | 0.4802 |
| **Earthquake** | 1.8447*** | 0.5033 | **0.0007** | 2.0919*** | 0.3841 | **0.0001** |
| **precision** | -0.0374 | 1.5812 | 0.9813 | -0.6998 | 0.6858 | 0.3134 |
| **Sample_size** | 0.3381 | 0.3284 | 0.3095 | 0.0003* | 0.0001 | **0.0225** |
| **China** | 0.0006** | 0.0002 | **0.0267** | -1.2488*** | 0.1322 | **0.0001** |
| **Year** | -1.41*** | 0.2203 | **0.0001** | -0.0663*** | 0.0113 | **0.0001** |
| **Netherlands** | -0.064*** | 0.0164 | **0.0003** | 0.442* | 0.2057 | **0.0375** |
| **Germany** | -0.5515* | 0.2982 | 0.0719 | -0.5004* | 0.2065 | 0.0198 |
| **Protest** | 0.2909 | 0.3121 | 0.357 | 0.1865 | 0.1176 | 0.1202 |
| **Random_sample** | 0.2091 | 0.1777 | 0.2462 | 0.7408*** | 0.1614 | **0.0001** |
| **Inverse_income** | 0.5211** | 0.1981 | **0.012** | 0.2799*** | 0.0637 | **0.0001** |
| **Age** | 0.3403*** | 0.0726 | **0.0001** | -0.0206** | 0.0072 | **0.0064** |
| **Top_ranked** | -0.0236* | 0.0128 | **0.0728** | -0.8085*** | 0.1346 | **0.0001** |
| **Low_citations** | -0.578*** | 0.1831 | **0.003** | -0.2988* | 0.1126 | **0.0112** |
| **Test moderatos** | | | | | | |
| F(df1 = 22, df2 = 42) | 12.3847 | | | 24.6629 | | |
| **p-val** | < .0001 | | | < .0001 | | |
| **sigma2 (level3)** | 0.0261 | | | 0.0000 | | |
| **sigma2.(level3/studies)** | 0.0272 | | | 0.0475 | | |

**Appendix I**

**W-RE**: weighted random effects model by the inverse of the standard error; **RE**: unweighted random effects model
*, **, and *** denote statistical significance at 10, 1%, and .1%, respectively.

**Table I.2**: Results of hierarchical model estimation (country as third level)

This table reports the meta-regression results using the random effects hierarchical method considering country as third level. The dependent variable of regression is the standardized willingness to pay (SWTP). Explanatory variables are defined in Table 3.

|  | W-RE | | | RE | | |
|---|---|---|---|---|---|---|
|  | Coeff. | SE | pval | Coeff. | se | pval |
| $\alpha_0$ (Precision) | -0.1957 | 0.6779 | 0.7742 | -0.1054 | 0.373 | 0.7789 |
| **Lab** | 1.423*** | 0.2135 | **0.0001** | 1.4585*** | 0.1867 | **0.0001** |
| **Online** | 0.2322 | 0.2072 | 0.2688 | 0.1844 | 0.1808 | 0.3135 |
| **Within_design** | 0.5433** | 0.1741 | **0.0033** | 0.5396*** | 0.1099 | **0.0001** |
| **Showup** | 1.437*** | 0.2509 | **0.0001** | 1.5404*** | 0.1885 | **0.0001** |
| **Incentive_compatible1** | -0.3602 | 0.2796 | 0.2047 | -0.3723. | 0.1983 | 0.0674 |
| **Elicit_method** | 0.0797 | 0.1067 | 0.4595 | 0.0892 | 0.0769 | 0.253 |
| **Probability_level** | -27.9665*** | 5.4187 | **0.0001** | -33.2259*** | 3.9707 | **0.0001** |
| **Implicit_prob** | -0.209 | 0.1645 | 0.211 | -0.2827* | 0.1157 | **0.0188** |
| **risk_idiosync** | 0.0395 | 0.2154 | 0.8553 | 0.1302 | 0.1828 | 0.4802 |
| **Earthquake** | 1.9783*** | 0.5103 | **0.0004** | 2.0919*** | 0.3841 | **0.0001** |
| **precision** | 0.127 | 1.2851 | 0.9217 | -0.6998 | 0.6858 | 0.3134 |
| **Sample_size** | 0.0005* | 0.0002 | **0.0154** | 0.0003* | 0.0001 | **0.0225** |
| **China** | -1.3885*** | 0.1659 | **0.0001** | -1.2488*** | 0.1322 | **0.0001** |
| **Year** | -0.0761*** | 0.0132 | **0.0001** | -0.0663*** | 0.0113 | **0.0001** |
| **Netherlands** | -0.6131* | 0.2794 | **0.0338** | 0.442* | 0.2057 | **0.0375** |
| **Germany** | 0.3067 | 0.3127 | 0.3324 | -0.5004* | 0.2065 | **0.0198** |
| **Protest** | 0.0735 | 0.1795 | 0.6842 | 0.1865 | 0.1176 | 0.1202 |
| **Random_sample** | 0.5957** | 0.1808 | **0.002** | 0.7408*** | 0.1614 | **0.0001** |
| **Inverse_income** | 0.3394*** | 0.0688 | **0.0001** | 0.2799*** | 0.0637 | **0.0001** |
| **Age** | -0.0187 | 0.0131 | 0.1597 | -0.0206** | 0.0072 | **0.0064** |
| **Top_ranked** | -0.6102** | 0.1782 | **0.0014** | -0.8085*** | 0.1346 | **0.0001** |
| **Low_citations** | -0.0735 | 0.1405 | 0.6035 | -0.2988* | 0.1126 | **0.0112** |
| Test moderatos $F(df1 = 22, df2 = 42)$ | | 18.7776 | | | 24.6629 | |
| **p-val** | | < .0001 | | | < .0001 | |
| **sigma2 (level3)** | | 0.0000 | | | 0.0000 | |
| **sigma2.(level3/studies)** | | 0.0475 | | | 0.0475 | |

**W-RE**: weighted random effects model by the inverse of the standard error; **RE**: unweighted random effects model
*, **, and *** denote statistical significance at 10, 1%, and .1%, respectively.

**Table I.3**: Results of hierarchical model estimation (survey-based data collection as third level)

This table reports the meta-regression results using the random effects hierarchical method considering survey-based data collection as third level. The dependent variable of regression is the standardized willingness to pay (SWTP). Explanatory variables are defined in Table 3.

|  | W-RE | | | RE | | |
|---|---|---|---|---|---|---|
|  | Coeff. | SE | pval | Coeff. | se | pval |
| $\alpha_0$ (Precision) | -0.1957 | 0.6779 | 0.7742 | -0.1054 | 0.373 | 0.7789 |
| Lab | 1.423*** | 0.2135 | 0.0001 | 1.4585*** | 0.1867 | 0.0001 |
| Online | 0.2322 | 0.2072 | 0.2688 | 0.1844 | 0.1808 | 0.3135 |
| Within_design | 0.5433** | 0.1741 | 0.0033 | 0.5396*** | 0.1099 | 0.0001 |
| Showup | 1.437*** | 0.2509 | 0.0001 | 1.5404*** | 0.1885 | 0.0001 |
| Incentive_compatible1 | -0.3602 | 0.2796 | 0.2047 | -0.3723. | 0.1983 | 0.0674 |
| Elicit_method | 0.0797 | 0.1067 | 0.4595 | 0.0892 | 0.0769 | 0.253 |
| Probability_level | -27.966*** | 5.4187 | 0.0001 | -33.225*** | 3.9707 | 0.0001 |
| Implicit_prob | -0.209 | 0.1645 | 0.211 | -0.2827* | 0.1157 | 0.0188 |
| risk_idiosync | 0.0395 | 0.2154 | 0.8553 | 0.1302 | 0.1828 | 0.4802 |
| Earthquake | 1.9783*** | 0.5103 | 0.0004 | 2.0919*** | 0.3841 | 0.0001 |
| precision | 0.127 | 1.2851 | 0.9217 | -0.6998 | 0.6858 | 0.3134 |
| Sample_size | 0.0005* | 0.0002 | 0.0154 | 0.0003* | 0.0001 | 0.0225 |
| China | -1.3885*** | 0.1659 | 0.0001 | -1.2488*** | 0.1322 | 0.0001 |
| Year | -0.0761*** | 0.0132 | 0.0001 | -0.0663*** | 0.0113 | 0.0001 |
| Netherlands | -0.6131* | 0.2794 | 0.0338 | 0.442* | 0.2057 | 0.0375 |
| Germany | 0.3067 | 0.3127 | 0.3324 | -0.5004* | 0.2065 | 0.0198 |
| Protest | 0.0735 | 0.1795 | 0.6842 | 0.1865 | 0.1176 | 0.1202 |
| Random_sample | 0.5957** | 0.1808 | 0.002 | 0.7408*** | 0.1614 | 0.0001 |
| Inverse_income | 0.3394*** | 0.0688 | 0.0001 | 0.2799*** | 0.0637 | 0.0001 |
| Age | -0.0187 | 0.0131 | 0.1597 | -0.0206** | 0.0072 | 0.0064 |
| Top_ranked | -0.6102** | 0.1782 | 0.0014 | -0.8085*** | 0.1346 | 0.0001 |
| Low_citations | -0.0735 | 0.1405 | 0.6035 | -0.2988* | 0.1126 | 0.0112 |
| Test moderatos | | | | | | |
| F(df1 = 22, df2 = 42) | | 18.7776 | | | 24.6629 | |
| p-val | | < .0001 | | | < .0001 | |
| sigma^2 (level3) | | 0.0000 | | | 0.0000 | |
| sigma^2.(level3/studies) | | 0.0475 | | | 0.0475 | |

**W-RE**: weighted random effects model by the inverse of the standard error; **RE**: unweighted random effects model
*, **, and *** denote statistical significance at 10, 1%, and .1%, respectively.

**Table I.4**: Results of hierarchical model estimation (Risk type as third level)

This table reports the meta-regression results using the random effects hierarchical method considering risk type as third level. The dependent variable of regression is the standardized willingness to pay (SWTP). Explanatory variables are defined in Table 3.

|  | W-RE | | | RE | | |
|---|---|---|---|---|---|---|
|  | Coeff. | SE | pval | Coeff. | se | pval |
| $\alpha_0$ (Precision) | -0.1957 | 0.6779 | 0.7742 | -0.1054 | 0.373 | 0.7789 |
| Lab | 1.423*** | 0.2135 | **0.0001** | 1.4585*** | 0.1867 | **0.0001** |
| Online | 0.2322 | 0.2072 | 0.2688 | 0.1844 | 0.1808 | 0.3135 |
| Within_design | 0.5433** | 0.1741 | **0.0033** | 0.5396*** | 0.1099 | **0.0001** |
| Showup | 1.437*** | 0.2509 | **0.0001** | 1.5404*** | 0.1885 | **0.0001** |
| Incentive_compatible1 | -0.3602 | 0.2796 | 0.2047 | -0.3723. | 0.1983 | 0.0674 |
| Elicit_method | 0.0797 | 0.1067 | 0.4595 | 0.0892 | 0.0769 | 0.253 |
| Probability_level | -27.9665*** | 5.4187 | **0.0001** | -33.2259*** | 3.9707 | **0.0001** |
| Implicit_prob | -0.209 | 0.1645 | 0.211 | -0.2827* | 0.1157 | **0.0188** |
| risk_idiosync | 0.0395 | 0.2154 | 0.8553 | 0.1302 | 0.1828 | 0.4802 |
| Earthquake | 1.9783*** | 0.5103 | **0.0004** | 2.0919*** | 0.3841 | **0.0001** |
| precision | 0.127 | 1.2851 | 0.9217 | -0.6998 | 0.6858 | 0.3134 |
| Sample_size | 0.0005* | 0.0002 | **0.0154** | 0.0003* | 0.0001 | **0.0225** |
| China | -1.3885*** | 0.1659 | **0.0001** | -1.2488*** | 0.1322 | **0.0001** |
| Year | -0.0761*** | 0.0132 | **0.0001** | -0.0663*** | 0.0113 | **0.0001** |
| Netherlands | -0.6131* | 0.2794 | **0.0338** | 0.442* | 0.2057 | **0.0375** |
| Germany | 0.3067 | 0.3127 | 0.3324 | -0.5004* | 0.2065 | **0.0198** |
| Protest | 0.0735 | 0.1795 | 0.6842 | 0.1865 | 0.1176 | 0.1202 |
| Random_sample | 0.5957** | 0.1808 | **0.002** | 0.7408*** | 0.1614 | **0.0001** |
| Inverse_income | 0.3394*** | 0.0688 | **0.0001** | 0.2799*** | 0.0637 | **0.0001** |
| Age | -0.0187 | 0.0131 | 0.1597 | -0.0206** | 0.0072 | **0.0064** |
| Top_ranked | -0.6102** | 0.1782 | **0.0014** | -0.8085*** | 0.1346 | **0.0001** |
| Low_citations | -0.0735 | 0.1405 | 0.6035 | -0.2988* | 0.1126 | **0.0112** |
| Test moderatos | | | | | | |
| F(df1 = 22, df2 = 42) | | 18.7776 | | | 24.6629 | |
| p-val | | < .0001 | | | < .0001 | |
| sigma2 (level3) | | 0.0000 | | | 0.0000 | |
| sigma2.(level3/studies) | | 0.0475 | | | 0.0475 | |

**W-RE**: weighted random effects model by the inverse of the standard error; **RE**: unweighted random effects model

**Table I.5**: Results of hierarchical model estimation (Coverage type as third level)

This table reports the meta-regression results using the random effects hierarchical method considering coverage type as third level. The dependent variable of regression is the standardized willingness to pay (SWTP). Explanatory variables are defined in Table 3.

|  | W-RE | | | RE | | |
|---|---|---|---|---|---|---|
|  | Coeff. | SE | pval | Coeff. | se | pval |
| $\alpha_0$ (Precision) | -0.1957 | 0.6779 | 0.7742 | -0.1054 | 0.373 | 0.7789 |
| Lab | 1.423*** | 0.2135 | 0.0001 | 1.4585*** | 0.1867 | 0.0001 |
| Online | 0.2322 | 0.2072 | 0.2688 | 0.1844 | 0.1808 | 0.3135 |
| Within_design | 0.5433** | 0.1741 | 0.0033 | 0.5396*** | 0.1099 | 0.0001 |
| Showup | 1.437*** | 0.2509 | 0.0001 | 1.5404*** | 0.1885 | 0.0001 |
| Incentive_compatible1 | -0.3602 | 0.2796 | 0.2047 | -0.3723. | 0.1983 | 0.0674 |
| Elicit_method | 0.0797 | 0.1067 | 0.4595 | 0.0892 | 0.0769 | 0.253 |
| Probability_level | -27.966*** | 5.4187 | 0.0001 | -33.225*** | 3.9707 | 0.0001 |
| Implicit_prob | -0.209 | 0.1645 | 0.211 | -0.2827* | 0.1157 | 0.0188 |
| risk_idiosync | 0.0395 | 0.2154 | 0.8553 | 0.1302 | 0.1828 | 0.4802 |
| Earthquake | 1.9783*** | 0.5103 | 0.0004 | 2.0919*** | 0.3841 | 0.0001 |
| precision | 0.127 | 1.2851 | 0.9217 | -0.6998 | 0.6858 | 0.3134 |
| Sample_size | 0.0005* | 0.0002 | 0.0154 | 0.0003* | 0.0001 | 0.0225 |
| China | -1.3885*** | 0.1659 | 0.0001 | -1.2488*** | 0.1322 | 0.0001 |
| Year | -0.0761*** | 0.0132 | 0.0001 | -0.0663*** | 0.0113 | 0.0001 |
| Netherlands | -0.6131* | 0.2794 | 0.0338 | 0.442* | 0.2057 | 0.0375 |
| Germany | 0.3067 | 0.3127 | 0.3324 | -0.5004* | 0.2065 | 0.0198 |
| Protest | 0.0735 | 0.1795 | 0.6842 | 0.1865 | 0.1176 | 0.1202 |
| Random_sample | 0.5957** | 0.1808 | 0.002 | 0.7408*** | 0.1614 | 0.0001 |
| Inverse_income | 0.3394*** | 0.0688 | 0.0001 | 0.2799*** | 0.0637 | 0.0001 |
| Age | -0.0187 | 0.0131 | 0.1597 | -0.0206** | 0.0072 | 0.0064 |
| Top_ranked | -0.6102** | 0.1782 | 0.0014 | -0.8085*** | 0.1346 | 0.0001 |
| Low_citations | -0.0735 | 0.1405 | 0.6035 | -0.2988* | 0.1126 | 0.0112 |
| Test moderatos F(df1 = 22, df2 = 42) | 18.7776 | | | 24.6629 | | |
| p-val | < .0001 | | | < .0001 | | |
| sigma2 (level3) | 0.0000 | | | 0.0000 | | |
| sigma2.(level3/studies) | 0.0475 | | | 0.0475 | | |

**W-RE**: weighted random effects model by the inverse of the standard error; **RE**: unweighted random effects model
*, **, and *** denote statistical significance at 10, 1%, and .1%, respectively.

**Appendix J: Repeated leave-one-out and meta-regression**

This table reports the meta-regression results after removing the three influential studies. The dependent variable of regression is the standardized willingness to pay (SWTP). Explanatory variables are defined in Table 3.

Table J.1 Estimation of the multivariate meta-regression

| | W-RE | | |
|---|---|---|---|
| | Coef. | SE | pval |
| $\alpha_0$ (Precision) | 0.2236 | 0.4804 | 0.6442 |
| Lab | 1.3138*** | 0.1415 | .0001 |
| Online | 0.0314 | 0.1474 | 0.8324 |
| Within_design | 0.3075* | 0.1233 | 0.017 |
| Showup | 1.2406*** | 0.1776 | .0001 |
| Incentive_compatible | -0.3058 | 0.1911 | 0.1176 |
| Elicit_method | 0.0986 | 0.0761 | 0.203 |
| Probability_level | -24.752*** | 3.9403 | .0001 |
| Implicit_prob | -0.2644* | 0.1005 | 0.0121 |
| Risk_idiosync | 0.1093 | 0.1479 | 0.4641 |
| Earthquake | 1.8137*** | 0.3399 | .0001 |
| Sample size | -2.0409* | 0.8806 | 0.0258 |
| China | 0.0004** | 0.0001 | 0.0064 |
| Year | -0.9999*** | 0.1119 | .0001 |
| Germany | -0.0574*** | 0.0091 | .0001 |
| Netherlands | 0.4043* | 0.2103 | 0.0618 |
| Protest | -0.2911 | 0.1986 | 0.1506 |
| Random_sample | 0.1994* | 0.1097 | 0.0768 |
| Inverse_income | 0.7248*** | 0.1328 | .0001 |
| Age | 0.226*** | 0.0491 | .0001 |
| Top_ranked | -0.0252* | 0.0094 | 0.0111 |
| Low_citations | -0.7126*** | 0.1262 | .0001 |
| R² | 95.60% | | |
| H² | 15.44 | | |
| Tau² (SE) | 0.0209 (0.0062) | | |
| I² | 93.53% | | |
| N-obs | 62 | | |

**Notes:** The results are from a cluster-random-effects model (3) a weighted random effects model by the inverse of the standard error (Weighted RE). All variables are described in Table 4. *, **, and *** denote statistical significance at 10, 1%, and .1%, respectively.

**Appendix K: Conditional WTP meta-regression**
This table reports the meta-regression results. The dependent variable is the logarithm of conditional WTP (non-zero WTP) divided by expected loss. Explanatory variables are defined in Table 3.

Table K.1 Estimation of the multivariate meta-regression

|  | W-RE | | |
| --- | --- | --- | --- |
|  | Coef. | SE | pval |
| $\alpha_0$ (Precision) | -1.4002** | 0.4957 | 0.0071 |
| Lab | 0.7785* | 0.4039 | 0.0605 |
| Online | -0.237 | 0.3616 | 0.5156 |
| Within_design | 0.672* | 0.2856 | 0.0233 |
| Showup | 1.788*** | 0.4535 | 0.0003 |
| Incentive_compatible | -1.1488** | 0.4177 | 0.0087 |
| Elicit_method | 0.5683* | 0.2247 | 0.0152 |
| Probability_level | -25.161* | 10.5566 | 0.0216 |
| Implicit_prob | -0.3449 | 0.2737 | 0.2144 |
| Risk_idiosync | 0.2033 | 0.3279 | 0.5386 |
| Earthquake | 3.2583** | 0.9871 | 0.0019 |
| Sample size | 4.5288* | 2.4181 | 0.0679 |
| China | 0.0005 | 0.0004 | 0.2567 |
| Year | -1.3931*** | 0.3389 | 0.0002 |
| Germany | -0.098*** | 0.0269 | 0.0007 |
| Netherlands | 1.3396** | 0.444 | 0.0043 |
| Protest | 0.101 | 0.434 | 0.8171 |
| Random_sample | -0.494 | 0.395 | 0.2178 |
| Inverse_income | 0.6732* | 0.3414 | 0.0551 |
| Age | 0.267* | 0.1422 | 0.0672 |
| Top_ranked | -0.8663* | 0.3371 | 0.0137 |
| Low_citations | -0.3422 | 0.2559 | 0.1881 |
| $R^2$ | 70.94% | | |
| $H^2$ | 155.96 | | |
| Tau² (SE) | 0.1817 (0.0411) | | |
| $I^2$ | 99.36% | | |
| N-obs | 65 | | |

Notes: The results are from a cluster-random-effects model (3) a weighted random effects model by the inverse of the standard error (Weighted RE). All variables are described in Table 4. *, **, and *** denote statistical significance at 10, 1%, and .1%, respectively.

## Appendix L:

**Table L1**: Estimation of the multivariate meta-regression without outliers removing

This table reports the meta-regression results. The dependent variable of regression is the standardized willingness to pay (SWTP). Explanatory variables are defined in Table 3. We consider a weighted random effects model by the inverse of the standard error (Weighted RE). The last column presents the regression results from BMA. For the BMA, the intercept posterior standard error is not available. As such, we recommend interpreting PIP value with caution.

|  | (1) Weighted RE | | | (2) Weighted BMA | | |
|---|---|---|---|---|---|---|
|  | Coef. | SE | pval | Post Mean | Post SE | PIP |
| $\alpha_0$ (Precision) | -0.2158 | -0.3983 | 0.696 | 0.0771 | 0.2739 | 0.2028 |
| $\alpha_1$ (Pub. bias) | **2.8099** | **1.0212** | **0.3234** | **2.4651** | **NA** | **1** |
| Lab | **1.6754** | **7.7445** | **0.0001** | **1.4704** | **0.2209** | **1** |
| Online | 0.2921 | 1.0816 | 0.2965 | 0.0460 | 0.1768 | 0.237 |
| Within_design | **0.4865** | **4.1825** | **0.0008** | **0.4148** | **0.1110** | **0.997** |
| Showup | **1.2962** | **3.6064** | **0.0026** | **1.2145** | **0.2000** | **0.999** |
| Incentive_compatible | 0.0363 | 0.1135 | 0.9111 | 0.0079 | 0.0764 | 0.1204 |
| Elicit_method | -0.1336 | -0.639 | 0.5324 | -0.0148 | 0.0563 | 0.1963 |
| Probability_level | **-41.5479** | **-4.1945** | **0.0008** | **-35.1958** | **7.4851** | **0.9996** |
| Implicit_prob | **-0.3411** | **-2.0737** | **0.0557** | **-0.1594** | **0.1393** | **0.6746** |
| Risk_idiosync | 0.0637 | 0.366 | 0.7195 | 0.0156 | 0.1032 | 0.1533 |
| Earthquake | **1.7338** | **3.4549** | **0.0035** | **1.7883** | **0.4276** | **0.9965** |
| Sample size | **0.0007** | **6.91** | **0.0001** | **0.0007** | **0.0001** | **1** |
| China | **-1.321** | **-6.3709** | **0.0001** | **-1.3209** | **0.1660** | **1** |
| Year | **-0.0719** | **-4.9511** | **0.0002** | **-0.0819** | **0.0098** | **1** |
| Germany | 0.2174 | 0.7488 | 0.4656 | 0.1536 | 0.2193 | 0.4302 |
| Netherlands | -0.6861 | -2.0873 | 0.0543 | -0.2423 | 0.3188 | 0.4861 |
| Protest | 0.3348 | 1.7084 | 0.1082 | **0.1740** | **0.1426** | **0.7066** |
| Random_sample | **0.8595** | **3.1552** | **0.0065** | **0.7761** | **0.2157** | **0.9890** |
| Inverse_income | **0.3514** | **5.4432** | **0.0001** | **0.3003** | **0.0946** | **0.9804** |
| Age | **-0.0253** | **-2.5025** | **0.0244** | **-0.0315** | **0.0069** | **0.9985** |
| Top_ranked | **-0.5415** | **-3.0808** | **0.0076** | **-0.4255** | **0.2087** | **0.9023** |
| Low_citations | -0.0021 | -0.0128 | 0.9899 | -0.0022 | 0.0441 | 0.1252 |
| $R^2$ | 84.23% | | | - | | |
| $H^2$ | 232.75 | | | - | | |
| Q stat. (p.value) | 1178.94(0.0649) | | | - | | |
| $Tau^2$ (SE) | 0.313(0.0121) | | | - | | |
| $I^2$ | 99.57% | | | - | | |
| N-obs | 74 | | | 74 | | |

**Notes:** Variables with PIP above 0.5 or significant are emphasized in bold. SD = standard deviation. SE = standard error. PIP= posterior inclusion probability. N.A. = not available. The last rows report regression $R^2$, $H^2$, Q, $Tau^2$ and $I^2$ statistics.

**Appendix M:** A "Dynamic" Meta-Analysis

Research on the demand for LPHI insurance represents a vast array of scientific articles dispersed across several databases and fields (finance, development economics, climatology, and earth sciences, etc.), which are not necessarily available to all scholars. Furthermore, the study of the demand for insurance against LPHI risks based on stated preferences has steadily increased over the last few years, making the use of meta-analysis as a quantitative research technique even more pertinent. It also implies that data must be regularly updated to reflect reality.

To this end, we propose an online survey web page (http://tinyurl.com/Wtp-lphi-insurance), under construction) to allow researchers working on insurance demand through WTP measurement to update the database data by entering information about their study. We hope that this dynamic meta-analysis will promote the consolidation and dissemination of knowledge on this topic and contribute to making science more open.

51